%% file: main.tex
\documentclass{article}
\usepackage[utf8]{inputenc}
\usepackage{verbatim}
\usepackage{listings}
\usepackage[a4paper, margin=0.9in]{geometry}
\usepackage{braket}
\usepackage{enumitem, kantlipsum}
\usepackage{graphicx}
\usepackage{caption}
\usepackage{subcaption}
\usepackage[dvipsnames,table,xcdraw]{xcolor}
\usepackage{multicol}
\usepackage{mathtools}

\usepackage{csquotes}
\usepackage{amsmath}

\usepackage{url}
\usepackage{hyperref}

\usepackage{qcircuit}
\usepackage{amsthm,amssymb}
\usepackage{fdsymbol}
\usepackage{appendix}

\usepackage{arydshln}

\let\OLDthebibliography\thebibliography
\renewcommand\thebibliography[1]{
  \OLDthebibliography{#1}
  \setlength{\parskip}{0pt}
}

\title{Quantum circuit design for universal distribution \\using a superposition of classical automata}
\author{Aritra Sarkar, Zaid Al-Ars, Koen Bertels\\
Department of Quantum \& Computer Engineering\\
Delft University of Technology, The Netherlands}
\date{April, 2021}

\begin{document}

\maketitle

\begin{abstract}

In this research, we present a quantum circuit design and implementation for a parallel universal linear bounded automata.
This circuit is able to accelerate the inference of algorithmic structures in data for discovering causal generative models.
The computation model is practically restricted in time and space resources.
A classical exhaustive enumeration of all possible programs on the automata is shown for a couple of example cases.
The precise quantum circuit design that allows executing a superposition of programs, along with a superposition of inputs as in the standard quantum Turing machine formulation, is presented.
% This is the first time a full mechanistic quantum automata is implemented in the circuit model of quantum computation.
This is the first time, a superposition of classical automata is implemented on the circuit model of quantum computation, having the corresponding mechanistic parts of a classical Turing machine.
The superposition of programs allows our model to be used for experimenting with the space of program-output behaviors in algorithmic information theory. 
Our implementations on OpenQL and Qiskit quantum programming language is copy-left and is publicly available on GitHub.
    
\end{abstract}


\input{text}

\newpage
\bibliographystyle{unsrt}
\bibliography{ref.bib}

\end{document}

%% file: text.tex
\newpage
\section{Introduction} \label{sec:introduction}

The phenomenal success of data-driven approaches like deep learning has ushered automation in many spheres of human society over the last decade.
However, such black-box optimization often fail to provide insights on the mechanism underlying the set of observations about the physical process under study.
This limitation in explainability are increasingly becoming crucial with automation in sectors like healthcare.
In contrast, symbolic approaches have been successfully used to model, study and understand the causal relationships in natural phenomena and datasets.
For example, in genomics, it finds applications~\cite{sarkar2021estimating} in meta-biology, phylogenetic tree analysis and protein-protein interaction mapping.
A better understanding of the algorithmic structures in DNA sequences would greatly advance personalized medication, drug discovery and synthetic biology.

Defining an algorithmic process via symbolic manipulations requires a computation model.
The Turing machine model of computation is the cornerstone of theoretical computer science.
This simple mechanistic automata has the expressive power to capture any algorithmic process.
Thus, it can be used for generic modeling and hypothesis comparison across various scientific disciplines.
% It has many applications specifically in algorithmic information theory.

The set of transformations a computation model can undergo and the resulting space of outputs is central to understanding the causal structure of the physical phenomena we intend to model.
Except for the trivial cases, this remains intractable on classical computers since the space of all possible transformations grows exponentially with the number of states and symbols of the automata.
Thus, in practice they are approximated by restricting the resources available to the computational model like time/cycles and space/memory.
Even with such restrictions, relatively simple automata have been explored using supercomputing clusters~\cite{soler2014calculating}.
These results have found various applications in genomics, psychology, network science, image processing, etc.

The quantum computation model provides distinctive advantages for specific algorithms using the laws of quantum mechanics.
Recently, a generic (gate-based) circuit model of quantum Turing machine was proposed~\cite{molina2019revisiting} based on cellular automata.
In this work, we propose an alternative model from a mechanistic perspective, with a \textit{detailed circuit implementation with realistic assumptions on run-time and qubit resources} for the tape memory.
We do away with the homogeneous local structure resulting in execution of many inactive unitary transforms, thereby improving the total number of executed operations.
Our model intuitively allows encoding a \textit{superposition of classical programs} and evaluating their evolution after a predetermined number of cycles.

In this research, we present the exact scalable circuit using standard quantum gates required to simulate a superposition of this resource-bounded stored-program automata.
With the recent thrust in the realization of quantum computing hardware, this is a promising direction with multifaceted application that can extend the applications of approximating algorithmic metrics by enumerating automata configurations.
The availability of better quantum processors would allow this algorithm to be readily ported on a quantum accelerator with a quantum computing stack~\cite{bertels2020quantum}.
We are motivated by applications in soft-computing, specifically estimating lower semi-computable metrics like algorithmic probability and Kolmogorov complexity, with application in genomics, artificial life and artificial intelligence.

% The necessary background of classical automata models and computational complexity of algorithms is presented in \S~\ref{s2} and \S~\ref{s3}.
The necessary background of classical automata models is presented in \S~\ref{s2}.
Thereafter, the quantum automata models and their circuit implementations are reviewed in \S~\ref{s4} .
In \S~\ref{s5} we propose our computation model and its key features that distinguish it from other approaches.
An exhaustive enumeration of a few small examples of the model is presented.
The quantum implementation of the QPULBA model is presented in \S~\ref{s6}.
The resources and circuit designs of each functional block of the quantum algorithm are worked out for one of the example cases of 2 state and 2 symbol automata.
In \S~\ref{s7} we present our results of the implementation of two cases of the QPULBA on two quantum programming platforms, OpenQL and Qiskit.
\S~\ref{s8} concludes the paper.

%%%%%%%%%%%%%%%%%%%%%%%%%%%%%%%%%%%%%%%%%%%%%%%%%%%
%%%%%%%%%%%%%%%%%%%%%%%%%%%%%%%%%%%%%%%%%%%%%%%%%%%
%%%%%%%%%%%%%%%%%%%%%%%%%%%%%%%%%%%%%%%%%%%%%%%%%%%
%%%%%%%%%%%%%%%%%%%%%%%%%%%%%%%%%%%%%%%%%%%%%%%%%%%
%%%%%%%%%%%%%%%%%%%%%%%%%%%%%%%%%%%%%%%%%%%%%%%%%%%
%%%%%%%%%%%%%%%%%%%%%%%%%%%%%%%%%%%%%%%%%%%%%%%%%%%
%%%%%%%%%%%%%%%%%%%%%%%%%%%%%%%%%%%%%%%%%%%%%%%%%%%
%%%%%%%%%%%%%%%%%%%%%%%%%%%%%%%%%%%%%%%%%%%%%%%%%%%

\newpage
\section{Classical automata models} \label{s2}

In this section, we briefly present the Turing machine model of computation and the variants of the model relevant for further discussions in the paper.
This is presented from the perspective of the Chomsky hierarchy of classical automata and languages.

\subsection{Turing machine model} \label{s2s1}

\begin{figure}[ht]
    \centering % \captionsetup{justification=centering} % trim: LBRT
    \includegraphics[clip, trim=19cm 0cm 0cm 12cm, width=0.6\textwidth]{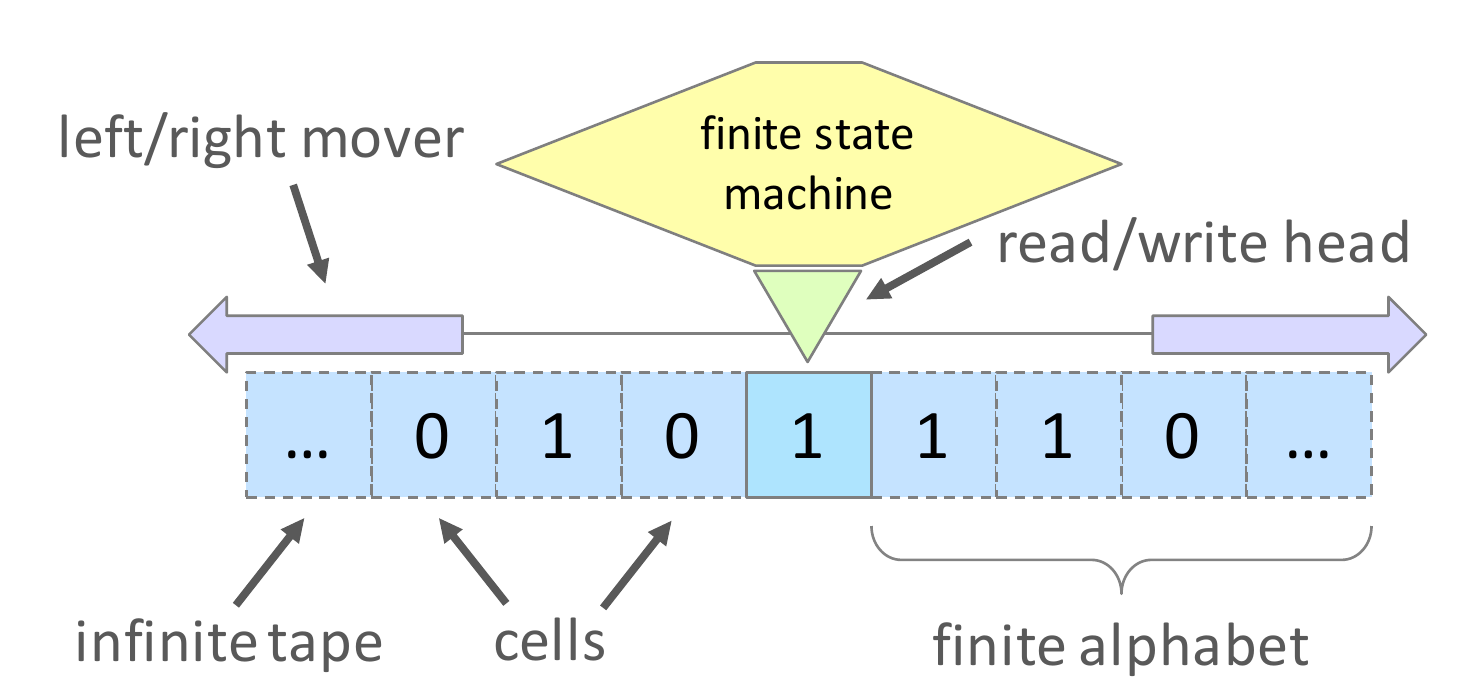}
    \caption{Computational model of a Turing machine}
    \label{fig:tm}
\end{figure}

The Turing machine (TM) is the canonical model of computation invented by Alan Turing for proving the uncomputability of the Entscheidungsproblem.
A TM manipulates symbols on an infinite memory strip of tape divided into discrete cells according to a table of transition rules.
These user-specified transition rules can be expressed as a finite state machine (FSM) which can be in one of a finite number of states at any given time and can transition between states in response to external inputs.
The Turing machine, as shown in figure~\ref{fig:tm}, positions its head over a cell and reads the symbol there.
As per the read symbol and its present state in the table of instructions, the machine (i) writes a symbol (e.g. a character from a finite alphabet) in the cell, then (ii) either moves the tape one cell left or right, and (iii) either proceeds to a subsequent instruction or halts the computation.

Each model of computation defines a set of inputs that are accepted by that automata.
For the Turing machine, this subset (of all possible strings over the alphabet of the language) that can be enumerated (outputs the string) is called recursively enumerable.
Recursively enumerable languages (also called, Turing recognizable/acceptable, semi/partially decidable) forms the type-0 in the Chomsky hierarchy of formal languages, as discussed in \S~2 \ref{s2s3}.

\subsection{Variants of the TM model} \label{s2s2}

While Turing machines can express arbitrary mechanical computations, their minimalist design makes them unsuitable for algorithm design for computation in practice.
Various modifications of the TM results in equivalent computation power, as captured by the Church-Turing thesis.
Here, equivalent refers to being within polynomial translation overhead in time or memory resources.
Thus, in practice they might compute faster, use less memory, or have smaller instruction set, but they cannot compute more mathematical functions.
These modified models of computation, like lambda calculus, Post machine, cyclic-tag system, offers different perspective in development of computer hardware and software.
Extending the TM tape to multiple dimensions is also equivalent to a 1 dimensional tape (or, only 1-way infinite tape).

In this section, we discuss some variants of the Turing machine model, as listed in Table~\ref{fig:TMs}, that are relevant for further ideas developed in this paper.
These variants can also be applied to the corresponding language instead of the automata.
Turing completeness is the ability for a system of instructions to simulate a Turing machine.
A Turing complete programming language is thus theoretically capable of expressing all tasks that can be accomplished by computers.
Nearly all programming languages are Turing complete if the limitations of finite memory are ignored.

\begin{figure}[htb]
    \centering % \captionsetup{justification=centering} % trim: LBRT
    \includegraphics[clip, trim=18cm 0cm 0cm 10cm, width=0.65\textwidth]{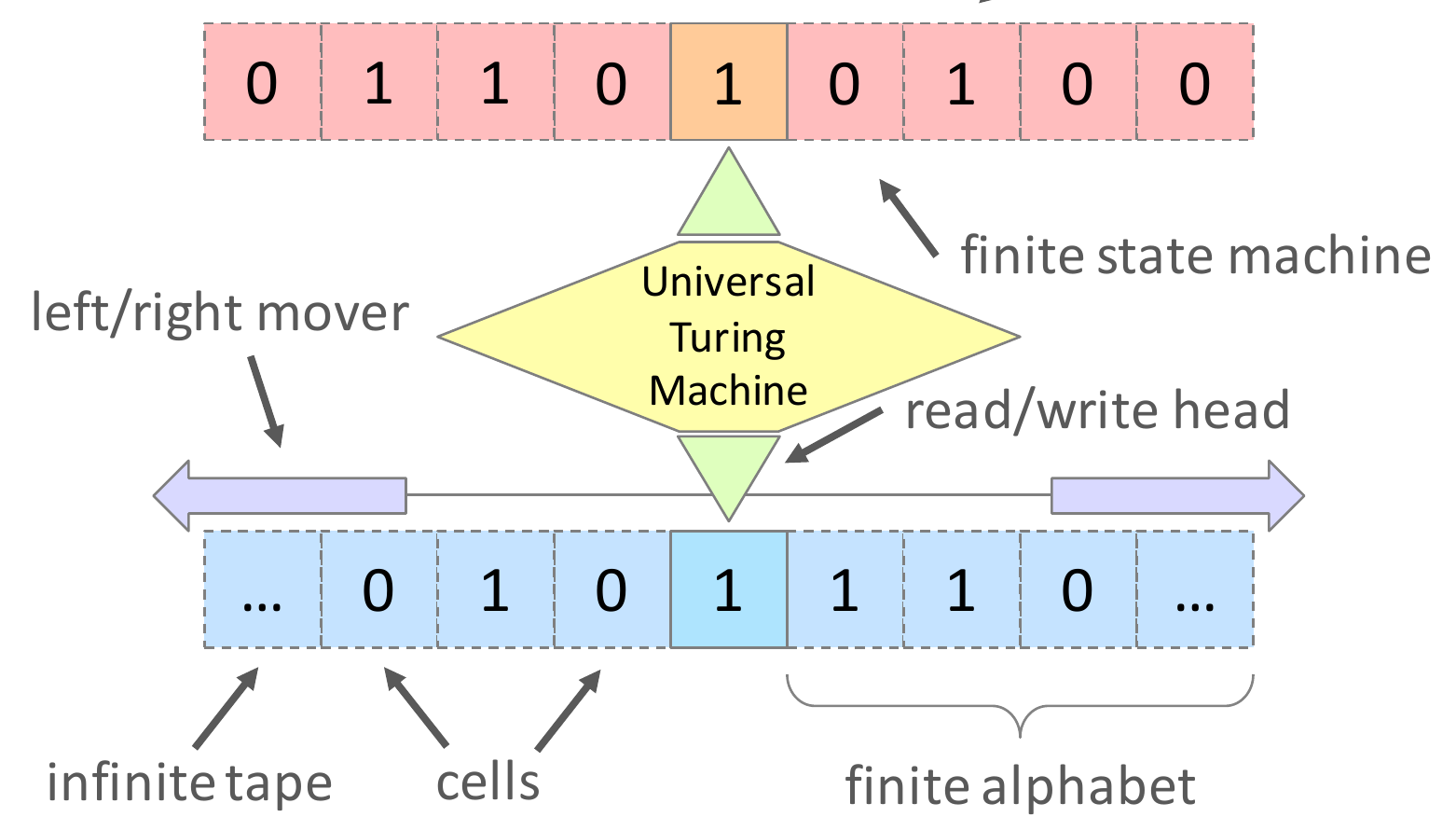}
    \caption{Computational model of a universal Turing machine}
    \label{fig:utm}
\end{figure}
A universal Turing machine (UTM) simulates an arbitrary Turing machine on an arbitrary input.
Every TM can be assigned a number, called the machine's description number, similar to Gödel numbers, encoding the FSM as a list of transitions.
The existence of this direct correspondence between natural numbers and TM implies that the set of all Turing machines (or programs of a fixed size) is denumerable.
A UTM reads the description of the machine to be simulated as well as the input to that machine from its own tape, as shown in figure~\ref{fig:utm}.
This is the idea behind the stored-program von Neumann architecture.
This is the first variation to automata models producing the automata variants in Table~\ref{fig:TMs}.

The sequential tape movement in TM/UTM makes them unsuitable for computation in practice (e.g. a binary search is really slow on a TM).
This is alleviated by extending the capability of the memory to access any indexed tape cell.
These variants of TM and UTM models are called random-access machine (RAM) and random-access stored-program (RASP) model respectively.
This is the second variation to automata models in Table~\ref{fig:TMs}.

Multiple FSMs can act based on the same tape data, allowing a shared memory for a multi-core processing model.
This modification is called the parallel TM (PTM) model and equivalently the PUTM, PRAM and PRASP, as the third variation to TM models in Table~\ref{fig:TMs}.

\begin{table}[p]
    \caption{Variants of the Turing machine model for deterministic, non-deterministic and quantum automata classes. Our implementation of QPULBA (in yellow) captures the computing capabilities of 27 (in blue) out of 51 automata models (of type 3, 2, 1) that is realistically implementable on physical hardware.}
    \label{fig:TMs}
    \centering % \captionsetup{justification=centering} % trim: LBRT
    \includegraphics[clip, trim=25.3cm 0cm 0cm 7.5cm, width=0.95\textwidth]{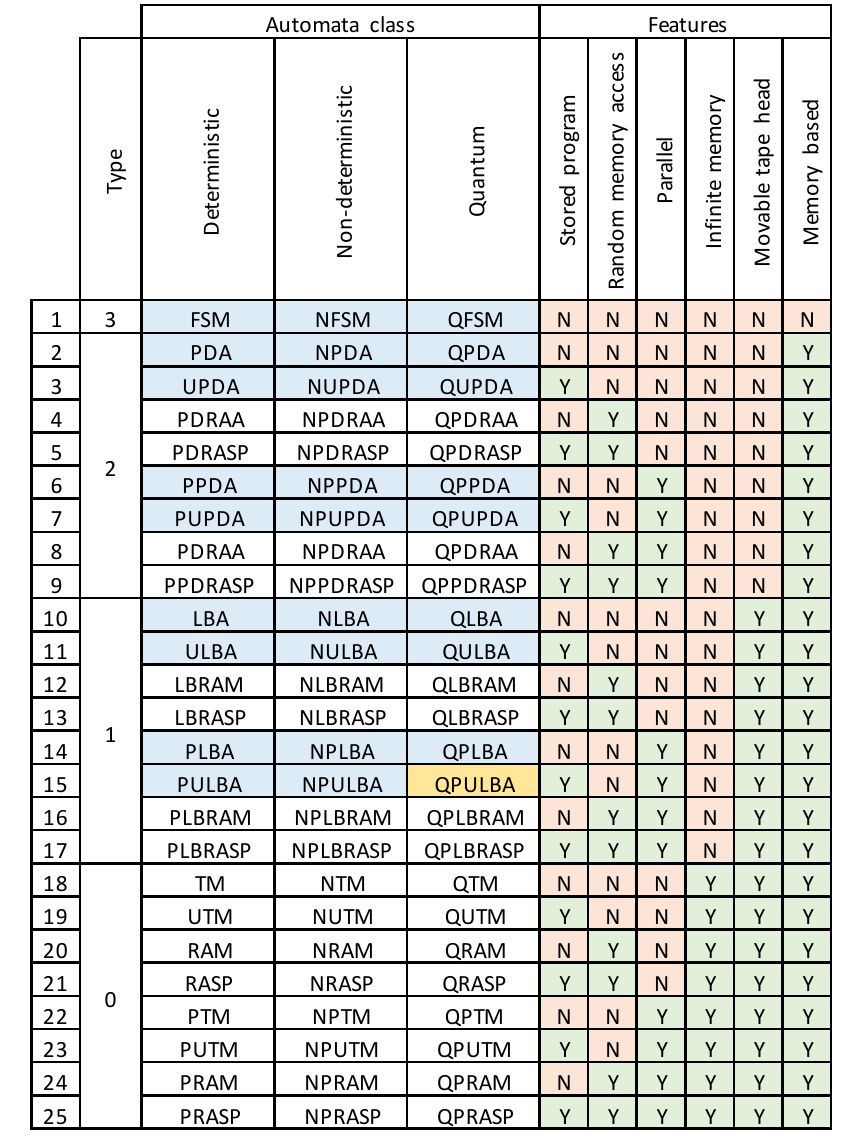}
\end{table}

\subsection{Chomsky hierarchy of automata and languages} \label{s2s3}

% A semi-weak and weak UTM applies to infinitely long tapes but with a repeated string (instead of a blank character) written in it 1-way or both ways.
For real implementations of an automata an infinite tape is not possible.
Thus computational models restrict the tape features in various ways~\cite{searls2002language,hauser2017universal}.
These result in a reduced level in the Chomsky hierarchy of formal languages and the corresponding automata model that accepts strings from the respective grammar.
\begin{itemize}[nolistsep,noitemsep]
    \item Type-0: Recursively enumerable language; Turing machine (TM)
    \item Type-1: Context-sensitive language; Linear bounded automata (LBA)
    \item Type-2: Context-free language; Pushdown automaton (PDA)
    \item Type-3: Regular languages; Finite state machine (FSM)
\end{itemize}
Each higher level (the highest being type-0) can always simulate the lower level.
Restrictions in memory makes the other levels computationally less capable than a TM in terms of the language, (i.e. the set of string patterns called the grammar) it can recognize.
For example, an FSM cannot determine if an input string has the structure $a^ib^i$, like $aabb,aaaabbbb$.
At type-0, universality is reached, thus, everything that is computable can be mapped to an algorithmic process on the TM.
This includes quantum computation as well, as it can be simulated on a classical TM (albeit using worst-case exponential resources).

% The discrete nature of the physical universe is a central thesis in the computational approach to science.
% Digital physicists support that there is a fundamental limit to dividing space and time - be it the discrete cells and steps of a Turing machine or that of Planck units.
% However, a lot of the power of the Turing machine relies on the infinite length of the tape.
% It is not too hard to construct logical paradoxes like Zeno's if we accept the concept of infinite but reject that of the infinitesimal.
A linear bounded automata (LBA) has a finite memory, extending the same modifications as for the TM, as listed as the fourth modification to TM models in Table~\ref{fig:TMs} forming the type-1 in the Chomsky hierarchy.
Most real-world computers can be best modeled as the Parallel Linear Bounded Random Access Stored Program (PLBRASP) automata.
This allows multiple concurrent processors with a shared random access finite sized memory.
The program as well as the data are accessed from the memory in the von Neumann architecture.

In this work, we will consider the Parallel Universal Linear Bounded Automata (PULBA) model which is the sequential memory access variant of the PLBRASP.
Thus, we consider multiple stored programs running in parallel while sharing a common limited work memory and output interface.

\subsection{Cellular automata model} \label{s2s4}

Often for physical system, a cellular automata model is preferred as an alternative to Turing machines, specifically when there is a homogeneous local interaction, e.g. in geographic spacial planning and ecological systems.
Cellular automata is a discrete model of computation consisting of a regular grid of cells in any finite dimension (or type of tessellation).
Each cell at any step in time is in one of a finite number of states, (such as on and off). 
For each cell, a set of cells called its neighborhood is defined relative to the specified cell. 
The next step in time is created according to some fixed rule (typically same over the entire space) that determines the new state of each cell in terms of the current state of the cell and the states of the cells in its neighborhood. 
The whole grid typically updates simultaneously.
The model with 1-dimensional tape over 2 symbols and a neighborhood size of 3 (1 cell in either size along with the current cell), is called elementary cellular automata (ECA).
These were studied extensively revealing interesting features like the Rule 110 (one of the 256 possible rules) capable of universal computation~\cite{wolfram2002new}.

\begin{figure}[htb]
    \centering % \captionsetup{justification=centering} % trim: LBRT
    \includegraphics[clip, trim=0cm 0cm 0cm 0cm, width=0.5\textwidth]{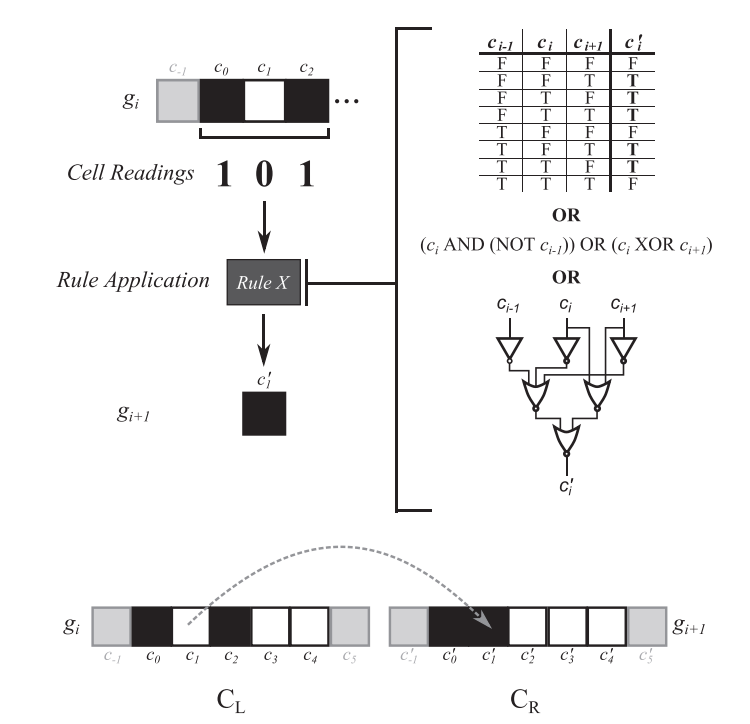}
    \caption{General execution of an elementary cellular automaton~\cite{chaplin2014photochromic}}
    \label{fig:sseca}
\end{figure}

Solid-state implementation~\cite{chaplin2014photochromic} of cellular automata involves expressing the update rule in terms of classical logic gates (e.g. AND, OR, NOT, NAND) as shown in figure~\ref{fig:sseca} for ECA Rule 110.
For each cell, the cell and its two immediate neighbors are read, their values entered into the transition function, and the cell updated with the new value for the subsequent generation. 
The transition function  can be represented as a truth table, a logic function or a logic circuit. 
The result from this function is then written to the corresponding cell in a duplicate array. 
In this example, transition functions are being calculated from $\text{C}_\text{L}$ and written sequentially to $\text{C}_\text{R}$. 
Once a generation is complete, the original array ($\text{C}_\text{L}$ here) is cleared, and the results of rule applications to cells in $\text{C}_\text{R}$ are written to $\text{C}_\text{L}$, like a ping-pong buffer.

A cellular automata can simulate a TM by storing a activation bit indicating the position of the tape head and the current state for each cell.
The transition function of the TM is encoded as the update rule depending only on the current cell.
Thus, it is like PTM with identical FSM, with only one of them active at each cycle.
This model is wasteful in terms of logic gate operations as only one of the cells is active yet all the other logic gates need to be executed none the less.

\newpage
\section{Quantum automata using quantum circuits} \label{s4}

Quantum computing uses the laws of quantum mechanics to a computational advantage.
These includes reversible unitary transformation of quantum states utilizing superposition and entanglement.
The algorithm design involves harnessing constructive interference between solution state's complex amplitudes and destructive interference between non-solutions before an irreversible projective measurement is performed on a chosen basis.
A good algorithm results in a high probability of reading out the classical state of an acceptable solution to the application.

A quantum mechanical model of Turing machines was described by Paul Benioff using Hamiltonian models~\cite{benioff1980computer,benioff1982quantum}.
A computationally equivalent model using quantum gates (inspired by classical Boolean logic gates) in a quantum circuit was proposed by David Deutsch~\cite{deutsch1985quantum,deutsch1989quantum,yao1993quantum}.
Due to the more intuitive nature of the circuit model it has become the standard for quantum algorithm development.
A generalized quantum Turing machine (GQTM)~\cite{iriyama2006generalized}, which contains QTM as a special case and includes non-unitary dynamics (irreversible transition functions) as well, allow the representation of quantum measurements without classical outcomes~\cite{perdrix2006classically}.
From a quantum computer architecture perspective, QTM was compared~\cite{wang2020quantum} to quantum random-access machine (QRAM) and quantum random-access stored-program (QRASP) model, proving their equivalence in bounded-error polynomial-time formulations using the Solvay-Kitaev theorem.
The QRASP considers that a single program is stored in classical registers, and thus treated as classical data, while the work memory can be in a superposition.
A QRASP model of a quantum computer can encode a program as quantum data, consequently generalizing the QRASP model to parallel quantum random-access stored-program machines (PQRASP) allowing a superposition of programs~\cite{ying2018reasoning}.

Here we review the circuit implementations of quantum finite automata (QFA)~\cite{dunlavey1998simulation,say2014quantum}, quantum equivalent automata models~\cite{tsai2020quantum}, and more extensively QTM~\cite{molina2019revisiting} which inspires our circuit architecture of a quantum parallel universal linear bounded automata (QPULBA).
Like its classical counterpart, a quantum Turing machine (QTM) is an abstract model capturing the power of quantum mechanical process for computation.
Any quantum algorithm can be expressed formally as a particular QTM.
% QTM generalizes the classical Turing machine such that the internal states of a classical TM are replaced by pure or mixed states in a Hilbert space.
% The transition function is replaced by a collection of unitary matrices that map the Hilbert space to itself.

\subsection{Quantum Turing machines} \label{s4s1}

A classical TM is extended to the complex vector space for a QTM.
For a three-tape QTM (one tape holding the input, a second tape holding intermediate calculation results, and a third tape holding output):
\begin{itemize}[nolistsep,noitemsep]
    \item The set of states is replaced by a Hilbert space.
    \item The tape alphabet symbols are replaced by a Hilbert space (usually a different Hilbert space than the set of states).
    \item The blank symbol corresponds to the zero-vector.
    \item The input and output symbols are usually taken as a discrete set, as in the classical system; thus, neither the input nor output to a quantum machine need be a quantum system itself.
    \item The transition function is a generalization of a transition monoid. It is a collection of unitary matrices that are automorphisms of the Hilbert space.
    \item The initial state may be either a mixed state or a pure state.
    \item The set of final or accepting states is a subspace of the Hilbert space.
\end{itemize}
It is important to realize that this is merely a sketch than a formal definition of a quantum Turing machine.
Some important details like how often a measurement is performed are not explicitly defined.
This is circumvented by formal proofs showing equivalence with an oblivious QTM, whose running time is a function of the size of the input (independent of the structure of the input).
However, how to practically translate that to a priori knowledge of the number of steps the computation needs to be executed before collapsing the superposition still needs further exploration~\cite{linden1998halting,guerrini2017quantum}.

% \subsection{TBD?} \label{s4s2}

% More extensive review: \cite{dunlavey1998simulation,say2014quantum,tsai2020quantum}

\subsection{Cellular automata inspired QTM circuit} \label{s4s3}

QTM was developed in the 1990s as a formal model to represent quantum computation.
Recently, \cite{molina2019revisiting,tsai2020quantum} revived this direction of research by presenting quantum circuit formulations and discussing applications of simulating quantum automata.

\begin{figure}[ht]
    \centering % \captionsetup{justification=centering} % trim: LBRT
    \includegraphics[clip, trim=0cm 0cm 0cm 0cm, width=0.8\textwidth]{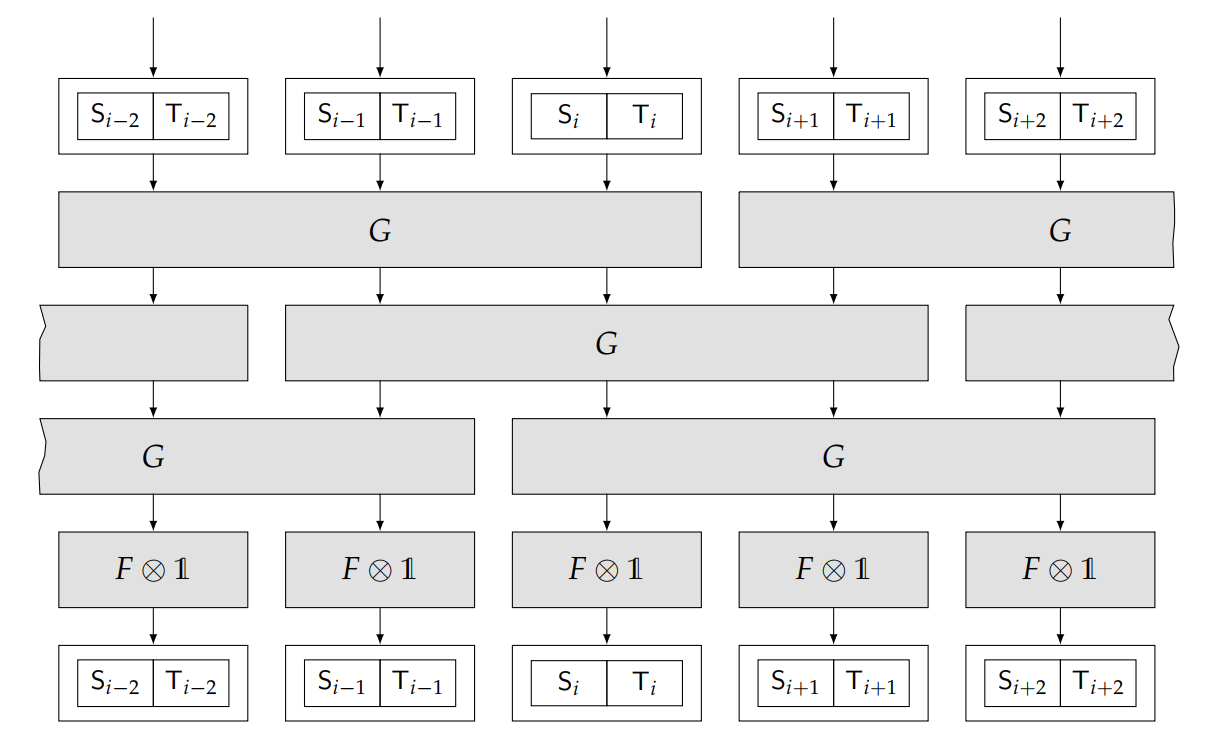}
    \caption{Universal computation in quantum CA models as quantum circuits~\cite{molina2019revisiting}}
    \label{fig:qeca}
\end{figure}

The approach taken in \cite{molina2019revisiting} resembles a (quantum) cellular automata in its homogeneous circuit template and local unitary operations.
The no-cloning principle prevents a direct translation of the sequentially updating classical cellular automata architecture to quantum circuits.
This is because, when the state is sequentially updated, we lose the neighbor information to update the consecutive cell.
The circuit architecture of this QTM is inspired by the solid-state implementation of the computational equivalent model to a TM of a 1-dimensional elementary cellular automata.
The local simultaneous update is achieved by maintaining a local one-hot marker denoting the presence/absence of the tape head at the location activates a specific (or superposition of) computation path in the QTM. 
It defines a fixed unitary $G$ which encodes the rule, or the transition function and interleaved in a particular way as shown in figure~\ref{fig:qeca}.
For the quantum version, the inputs are qubits, so they can be in a superposition of states, allowing simultaneous evolution of various inputs (initial Turing tapes).

By this approach, $t \ge n$ steps of a QTM on an input of length $n$ can be simulated by a uniformly generated family of quantum circuits linear in $t$.
However, not all transition functions describe valid quantum Turing machines as it is imperative that the global evolution is unitary on the Hilbert space.
Any QTM that exhibits a local causal behavior can be mapped to the transition function $G$ proposed in \cite{molina2019revisiting}.
Since the model is universal, there exists a transition function that will in effect read a stored program and the input from the tape as qubits and perform the computation entirely by local operations.
However, it is not immediately clear how to construct such a unitary for an arbitrary transition function.
Moreover, modeling the stored-program model on this local interaction architecture would incur a high cost in the number of operations to affect the tape, as at each step a TM head moves only by one step.

%%%%%%%%%%%%%%%%%%%%%%%%%%%%%%%%%%%%%%%%%%%%%%%%%%%
%%%%%%%%%%%%%%%%%%%%%%%%%%%%%%%%%%%%%%%%%%%%%%%%%%%
%%%%%%%%%%%%%%%%%%%%%%%%%%%%%%%%%%%%%%%%%%%%%%%%%%%
%%%%%%%%%%%%%%%%%%%%%%%%%%%%%%%%%%%%%%%%%%%%%%%%%%%
%%%%%%%%%%%%%%%%%%%%%%%%%%%%%%%%%%%%%%%%%%%%%%%%%%%
%%%%%%%%%%%%%%%%%%%%%%%%%%%%%%%%%%%%%%%%%%%%%%%%%%%
%%%%%%%%%%%%%%%%%%%%%%%%%%%%%%%%%%%%%%%%%%%%%%%%%%%
%%%%%%%%%%%%%%%%%%%%%%%%%%%%%%%%%%%%%%%%%%%%%%%%%%%
\newpage
\section{Proposed computation model} \label{s5}

In this section, we propose the formulation of our computation model that we translate to the quantum circuit in \S~\ref{s6}.
Our motivation is to complement the research as presented in \cite{molina2019revisiting} from a different perspective.
The major considerations to make the quantum circuit practically implementable and tractable in resources is discussed.

\subsection{A mechanistic perspective} \label{s5s1}

The intricacy in applying the cellular automata based QTM to arbitrary algorithms inspires our research to look at more intuitive model with structural similarity to a TM.
We do away with the local transition function applied in parallel based on activation qubits.
Our quantum circuit provides a mechanistic model of a Linear Bounded Automata, while allowing both the inputs (tape memory) and the program (transition function) to be in superposition.
Thus, in effect it provides a scalable quantum circuit implementation for the QPULBA model of computation.

The superposition of programs is the key feature allowing the framework to be applied for estimating algorithmic features~\cite{sarkar2021}, like the algorithmic probability and algorithmic complexity.
Note that, this parallelism we propose is different from the superposition of inputs as considered by standard QTM models \cite{molina2019revisiting}, as the parallelism we propose allows both the input data as well as the transition function to be in a superposition.
As an analogy, this can be thought of as the distinction between a quantum adder (which can add a superposition of two inputs), and a quantum calculator (which can apply all possible binary operations like add, subtract, multiply, divide, power, etc. on the two inputs).
Each classical binary operation need not access the inputs from the tape in a coordinated fashion, e.g. multiplication can proceed from right to left while division can proceed from left to right.
Once the quantum algorithm is executed, the output evolves to a superposition considering all possible programs (with the given states and symbols) on all possible inputs - commonly referred to as the Solomonoff universal prior probability distribution in algorithmic information theory.

\subsection{Bounds on runtime} \label{s5s2}

For any pragmatic (quantum) automata implementation, besides the memory, the runtime also needs to be restricted to a predetermined cycle count.
This is imperative as it is not possible to estimate the halting time of a Turing machine in advance.
Not all applications however allow a time restricted formulation.
In our research, we are interested in these specific types of algorithms which can be aborted externally (with or without inspecting the current progress).
The premature output yields an acceptable approximate solution with a bounded error based on the computation time.
They come under the purview of soft-computing and are ubiquitous in optimization, machine learning and evolutionary algorithms.

Here we consider runtime restriction in the context of a linear-bounded automata.
There are 2 variables in a LBA specification, apart from the alphabet and states.
These are the length of computation (time) and the length of the tape (space) resources.
The latter is the dependent variable due to the region of influence (time cone).
\textit{It is possible for a code to run for infinite time on a limited tape, but it is not possible to cover an infinite tape in a finite time}.
Thus, time-restricted automata further restrict the set of strings that are reachable by the computation process.
There is no definitive metric to compare the power of runtime restricted automata models without invoking algorithmic information metrics like the speed prior or logical depth.
Most landmark research~\cite{neary2009small} on the universality of TMs has assumed infinite tape length, adding a few points on the pareto curve of universality for the number of symbols and states.
The runtime of our model is thus application and resource driven, based on the convergence of the model for the specific phenomena, and the coherence limits of the quantum processor.

Additionally, we do not consider halting states specifically in our model.
However, it is not too difficult to consider the subset of programs with halting states and are by default realized when a state loops on itself.
Recent research~\cite{zenil2019coding} has shown that a non-halting TM is statistically correlated in rank with LBA.
Though the Chomsky hierarchy level of a non-halting LBA has not been explored, we presume that it would also be correlated to LBA rather than to PDA or FSM.
As discussed later, restricting the automata based on the number of cycles is an unavoidable technicality in the quantum version as we need to collapse the superposition.

% For restricted TM, instead of universality, more important metrics are expressibility and reachability.
% Expressibility is the set of solutions/states that an automaton can compute given a non-universal set of operations but not restricted by memory/space or steps/time resources.
% Reachability is the set of solutions/states that an automaton can compute by restricting space and/or time resources.
% For our computational model of a time-bound LBA, the reachability metric will be more suitable, semantically being the set of non-zero probability strings in the universal distribution we empirically obtain.

\subsection{Corresponding classical model} \label{s5s3}

Before we present the quantum model of QPULBA, in this section, we define the corresponding classical model where neither the program nor the input can be in a superposition.
Thus, it corresponds to a restricted runtime ULBA, where the program is accepted from outside the system (e.g. stored in a program memory) and the input tape is finite in length.

In our computation model, we will restrict a $m$ states, $n$ symbols, $1$ dimension tape Turing machine by limiting the maximum time steps $t$ before a forced halt is imposed.
This automatically bounds the causal cone on the tape to $[-t,+t]$ from the initial position of the tape head, resulting in a LBA.

The tape is initialized to the blank character as this does not reduce the computational power of the TM.
This can be thought of as, the initial part of the program prepares the input on the tape and then computes on it.
The tape length, like the memory size of CPU, is an application specific hyperparameter chosen such that it is enough for accommodating the intermediate work-memory scratchpad and the final output.
The range of values for the tape length is $c \le (2t+1)$.

The generic model of a restricted ULBA is represented by this 10-tuple.
\begin{align*} 
T &=  \braket{Q,\Gamma,b,\Sigma,\delta,q_0,F,t,d,c}
\end{align*}

% \newpage
\begin{itemize}[nolistsep,noitemsep]
    \item $Q$ is a finite, non-empty set of states.
    \item $\Gamma$ is a finite, non-empty set of symbols allowed on the tape, called the tape alphabet.
    \item $b\in \Gamma$ is the blank symbol.
    % It is the only symbol allowed to occur on the tape infinitely often at any step during the computation.
    \item $\Sigma \subseteq \Gamma \setminus \{b\}$ is the set of input symbols, that is, the set of symbols allowed to appear in the initial tape contents.
    \item $\delta$ is a partial function called the transition function. It defines the next state, tape movement and write symbol based on the current state and the read symbol.
    \item $q_0\in Q$ is the initial state.
    \item $F\subseteq Q$ is the set of final states or accepting states. 
    % The initial tape contents is said to be accepted by $T$ if it eventually halts in a state from $F$.
    \item $t$ is the number of steps the machine is executed before a forced halt is imposed.
    \item $d$ is the dimension of the tape. It also specifies if the tape is circular by a $\circ$ symbol for each dimension.
    \item $c$ is the length of the tape in each dimension.
\end{itemize}

For our experimental implementation example in this paper, we chose $c = t$ and a circular tape.
This helps us evaluate output diversity considering tape direction equivalence under left-right substitution.
This shorter tape however comes as the cost of not able to map to computational path dependent application where left-right substitution is not always equivalent (e.g. robotic movement).
Our computation model of $m = |Q|$ states, $n = |\Gamma|$ symbols, $d=1^\circ$ dimension (circular) tape restricted Turing machine, can be formally represented as:
\begin{align*} 
T &=  \braket{\{Q_0,Q_1,\dots,Q_{m-1}\},\{s_0,s_1,\dots,s_{n-1}\},s_0,\{\},\delta,Q_0,\{\},c,1^\circ,c}
\end{align*}
Note that $\Sigma$ is empty, thus the tape is always initialized to the blank character $b = s_0$.
The set of accepting states $F$ is also empty to prevent the machine from halting before $t$ steps.
This includes machines that have defined halting states, by modifying the transition function to remain in the same state and write the same symbol that is read (in effect simulating a no-operation) once these states are reached.

Thus, the transition table exhaustively lists a transition for each combination of $(Q-F) = Q$ and $\Gamma$.
$$\delta: Q\times \Gamma \rightarrow Q\times \Gamma \times \{0,1\}$$
where, $0$ is left shift, $1$ is right shift for the $1$ dimensional tape.
Each entry of the transition table requires $log_2(m)+log_2(n)+d$ bits and there are a total of $m*n$ entries.
Thus, the number of bits required to specify a single transition function is:
$$q_\delta = (m*n)*(log_2(m)+log_2(n)+d)$$
The number of different machines (programs) that can be represented using $q_\delta$ bits is represented by: 
$$P = 2^{q_\delta}$$

\subsection{Enumeration of example cases} \label{s5s4}

Here, some examples of this computation model are enumerated.
These enumerations are intended to be executed in parallel thereby presenting the PULBA variant of the ULBA model.
The cases are labeled as per the number of states, symbols and tape dimension (Case $m$-$n$-$d$).
We start with the smallest natural number $1$ and explore larger values of symbols and states (however, we will only consider $1$ dimensional Turing tape).

For these experiments, the number of iterations is set equal to the size of the program, i.e. $t = q_\delta$.
This allows us to compare the final tape and program to infer which programs can self-replicate, which motivates our research~\cite{sarkar2021}.
% The model as shown in figure~\ref{fig:rtm} is executed for $t$ iterations.
% \begin{figure}[ht]
%     \centering % \captionsetup{justification=centering} % trim: LBRT
%     \includegraphics[clip, trim=18cm 0cm 0cm 10cm, width=0.7\textwidth]{figures/rtm.pdf}
%     \caption{Computational model used for experimentation}
%     \label{fig:rtm}
% \end{figure}

\subsubsection{Case 1-1-1} \label{s5s4s1}
For this case, the number of states $m = 1$ with the state set $Q: \{Q_0\}$.
The alphabet is $\Gamma: \{0\}$, thus, $n = 1$ (the unary alphabet).
This gives the values $q_\delta = 1*1*(0+0+1) = 1$ and $P = 2^1 = 2$ using the formula discussed before.

The machine is run for $t = q_\delta = 1$ iteration.
The initial tape is of length $c = t = 1$.
To distinguish the blank character from a written $0$, we will use $o$.
Thus, the initial tape is: $o$.

The outputs for each program (description number) are listed in Table~\ref{tab:utm111}.
$R_s$/$W_s$ refers to the read/write symbols and $M_{l/r}$ refers to the tape movement of left/right.

\begin{table}[ht]
    \caption{Exhaustive enumeration of the programs of Case 1-1-1 circular tape ULBA for length 1 cycle 1}
    \label{tab:utm111}
    \centering
    \begin{tabular}{c|c|c}
        P\# & $Q_0 R_0$ & Final tape \\
        \hline
        0 & $Q_0 M_l W_0$ & \texttt{0} \\
        1 & $Q_0 M_r W_0$ & \texttt{0} \\
        \hline
    \end{tabular}
\end{table}

\subsubsection{Case 2-1-1} \label{s5s4s2}
For this case, the number of states $m = 2$ with the state set $Q: \{Q_0,Q_1\}$.
The alphabet is again $\Gamma: \{0\}$, with, $n = 1$ (the unary alphabet).
This gives the values $q_\delta = 2*1*(1+0+1) = 4$ and $P = 2^4 = 16$.

The machine is run for $t = q_\delta = 4$ iteration.
The initial tape is of length $c = t = 4$.
The initial tape is: $\underline{o}ooo$, where the underline denotes the initial position.

The outputs for each program (description number) are listed in Table~\ref{tab:utm211}.
% If the machine tries to read/write outside the tape index, the error flag prematurely ends the computation.
% For example, for machine $11$ we have $0ooo$ instead of $0000$.
It is easy to see all m-1-d cases will result in tapes with the blank/zero characters.
The $0000$ string is the only string in this language.
% Thus, only the description number $0$ case can be considered self-replicating.

\begin{table}[ht]
    \caption{Exhaustive enumeration of the programs of Case 2-1-1 circular tape ULBA for length 4 cycle 4}
    \label{tab:utm211}
    \centering
    \begin{tabular}{c|c:c|c}
        P\# & $Q_1 R_0$ & $Q_0 R_0$ & Final tape \\
        \hline
        00 & $Q_0 M_l W_0$ & $Q_0 M_l W_0$ &  \texttt{0000} \\
        01 & $Q_0 M_l W_0$ & $Q_0 M_r W_0$ &  \texttt{0000} \\
        02 & $Q_0 M_l W_0$ & $Q_1 M_l W_0$ &  \texttt{0000} \\
        03 & $Q_0 M_l W_0$ & $Q_1 M_r W_0$ &  \texttt{00oo} \\
		\hdashline
        04 & $Q_0 M_r W_0$ & $Q_0 M_l W_0$ &  \texttt{0000} \\
        05 & $Q_0 M_r W_0$ & $Q_0 M_r W_0$ &  \texttt{0000} \\
        06 & $Q_0 M_r W_0$ & $Q_1 M_l W_0$ &  \texttt{0oo0} \\
        07 & $Q_0 M_r W_0$ & $Q_1 M_r W_0$ &  \texttt{0000} \\
        \hdashline
		08 & $Q_1 M_l W_0$ & $Q_0 M_l W_0$ &  \texttt{0000} \\
        09 & $Q_1 M_l W_0$ & $Q_0 M_r W_0$ &  \texttt{0000} \\
        10 & $Q_1 M_l W_0$ & $Q_1 M_l W_0$ &  \texttt{0000} \\
        11 & $Q_1 M_l W_0$ & $Q_1 M_r W_0$ &  \texttt{00o0} \\
        \hdashline
		12 & $Q_1 M_r W_0$ & $Q_0 M_l W_0$ &  \texttt{0000} \\
        13 & $Q_1 M_r W_0$ & $Q_0 M_r W_0$ &  \texttt{0000} \\
        14 & $Q_1 M_r W_0$ & $Q_1 M_l W_0$ &  \texttt{00o0} \\
        15 & $Q_1 M_r W_0$ & $Q_1 M_r W_0$ &  \texttt{0000} \\
        \hline
    \end{tabular}
\end{table}

\subsubsection{Case 1-2-1} \label{s5s4s3}
For this case, the number of states $m = 1$ with the state set $Q: \{Q_0\}$.
The alphabet is $\Gamma: \{0,1\}$, with, $n = 2$ (the binary alphabet).
This gives the values $q_\delta = 1*2*(1+1+0) = 4$ and $P = 2^4 = 16$.
The machine is run for $t = q_\delta = 4$ iteration.
The initial tape is of length $c = t = 4$.
The initial tape is similar to the last case: $\underline{o}ooo$.

The outputs for each program (description number) are listed in Table~\ref{tab:utm121}.
% Machines $00$ and $15$ are self-replicating.
% This depends on how the state machine is encoded as the description number and where is the starting position of the tape.
% Strings $0000$ and $1111$ have a probability of $0.5$ each.
% Note that the transitions for $R_1$ are never effectively used as the head never returns to a cell which was previously written by $1$.
If we run the experiment for more than $4$ steps then we will see more variety in the output.

\begin{table}[ht]
    \caption{Exhaustive enumeration of the programs of Case 1-2-1 circular tape ULBA for length 4 cycle 4}
    \label{tab:utm121}
    \centering
    \begin{tabular}{c|c:c|c}
        P\# & $Q_0 R_1$ & $Q_0 R_0$ & Final tape \\
        \hline
        00 & $Q_0 M_l W_0$ & $Q_0 M_l W_0$ & \texttt{0000} \\
        01 & $Q_0 M_l W_0$ & $Q_0 M_l W_1$ & \texttt{1111} \\
        02 & $Q_0 M_l W_0$ & $Q_0 M_r W_0$ & \texttt{0000} \\
        03 & $Q_0 M_l W_0$ & $Q_0 M_r W_1$ & \texttt{1111} \\
        \hdashline
		04 & $Q_0 M_l W_1$ & $Q_0 M_l W_0$ & \texttt{0000} \\
        05 & $Q_0 M_l W_1$ & $Q_0 M_l W_1$ & \texttt{1111} \\
        06 & $Q_0 M_l W_1$ & $Q_0 M_r W_0$ & \texttt{0000} \\
        07 & $Q_0 M_l W_1$ & $Q_0 M_r W_1$ & \texttt{1111} \\
        \hdashline
		08 & $Q_0 M_r W_0$ & $Q_0 M_l W_0$ & \texttt{0000} \\
        09 & $Q_0 M_r W_0$ & $Q_0 M_l W_1$ & \texttt{1111} \\
        10 & $Q_0 M_r W_0$ & $Q_0 M_r W_0$ & \texttt{0000} \\
        11 & $Q_0 M_r W_0$ & $Q_0 M_r W_1$ & \texttt{1111} \\
        \hdashline
		12 & $Q_0 M_r W_1$ & $Q_0 M_l W_0$ & \texttt{0000} \\
        13 & $Q_0 M_r W_1$ & $Q_0 M_l W_1$ & \texttt{1111} \\
        14 & $Q_0 M_r W_1$ & $Q_0 M_r W_0$ & \texttt{0000} \\
        15 & $Q_0 M_r W_1$ & $Q_0 M_r W_1$ & \texttt{1111} \\
        \hline
    \end{tabular}
\end{table}

\subsubsection{Case 2-4-1} \label{s5s4s4}
Let us consider the case where the number of states $m = 2$ (with the states set $Q: \{Q_0,Q_1\}$) with the alphabet $\Gamma: \{A,C,G,T\}$, with, $n = 4$, inspired by the DNA alphabet.
This gives the values $q_\delta = 2*4*(1+2+1) = 32$ and $P = 2^{32} = 4294967296$.
It is clear that even for the simple case of the DNA alphabet, an exhaustive search by enumeration is not possible on a classical algorithm (and maybe even on a quantum computer simulator running on classical hardware).
% \subsubsection{Case 4-4-1} \label{s5s4s5}
% If we consider the number of states $m = 4$ (with the states set $Q: \{Q_A,Q_C,Q_G,Q_T\}$) corresponding to the alphabet $\Gamma: \{A,C,G,T\}$, with, $n = 4$ (the DNA alphabet), we can model the execution of the genetic code as a Markov process.
% This gives the values $q_\delta = 4*4*(2+2+1) = 80$ and $P = 2^{80} = 1208925819614629174706176$.
This exponential growth in the number of cases shows the difficulty of classical enumeration of ULBA motivating quantum acceleration~\cite{sarkar2019algorithm,sarkar2020quaser} for bioinformatics applications.

\subsubsection{Case 2-2-1} \label{s5s4s6}
This case is both non-trivial as well as within the bounds of our current experimentation.
The number of states $m = 2$ with the state set $Q: \{Q_0,Q_1\}$.
The alphabet is $\Gamma: \{0,1\}$, thus, $n = 2$ (the binary alphabet).
This gives the values $q_\delta = 2*2*(1+1+1) = 12$ and $P = 2^{12} = 4096$ using the formula discussed before.

The machine is run for $t = q_\delta = 12$ iteration.
The initial tape of length $c = t = 12$ is: $\underline{o}ooooooooooo$

The program (description number) is encoded as:
$$[QMW]^{Q_1R_1}[QMW]^{Q_1R_0}[QMW]^{Q_0R_1}[QMW]^{Q_0R_0}$$
There are too many cases to enumerate by hand, so a Python script (the classical kernel we intend to accelerate) is written that emulates our restricted model of the linear bounded automata for all 4096 cases.
The program can be found at the following link:
\\
\href{https://github.com/Advanced-Research-Centre/QuBio/blob/master/Project_01/classical/}{https://github.com/Advanced-Research-Centre/QuBio/blob/master/Project\_01/classical/}

\begin{figure}[ht]
\centering
\includegraphics[clip, trim=4cm 1cm 4cm 2cm, width=0.8\textwidth]{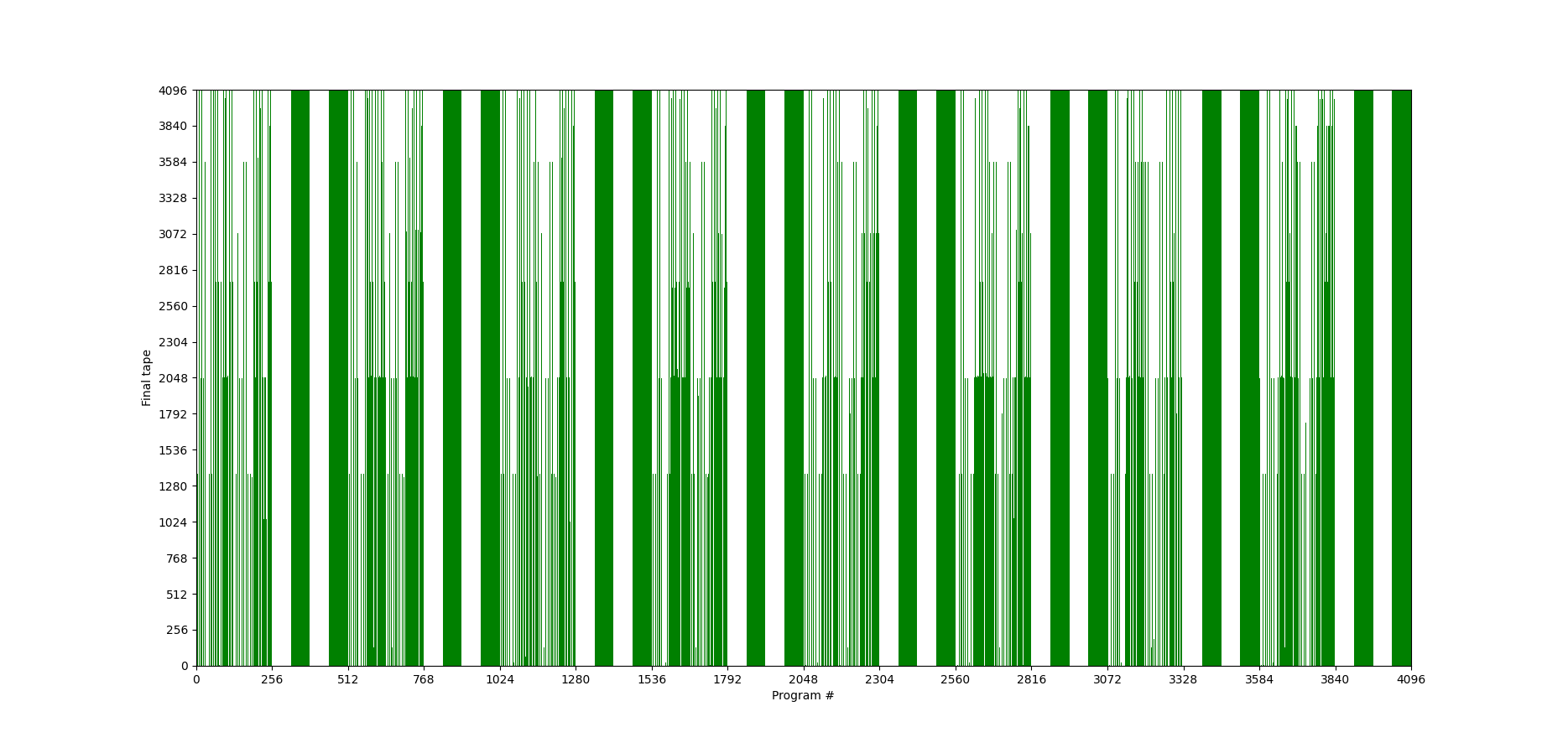}
\caption{Tape output for all programs of Case 2-2-1 circular tape ULBA for length 12 cycle 12}
\label{fig:utm221}
\end{figure}

% We find machines $0$ and $4095$ are self-replicating.
The tape output for all the 4096 machines is plotted in figure~\ref{fig:utm221}.

\newpage
\section{QPULBA: quantum parallel universal linear bounded automata} \label{s6}

\begin{figure}[htb]
\centering 
\centerline{
\Qcircuit @C=1em @R=1.2em {
\lstick{\ket{q_{a}}}	 	& \qw 	& \multigate{8}{Init} 	& \qw 	&  \pushs{\rule{3.4em}{0.01em}} \qw 	& \qw 								& \qw 					& \gate{Move} \qwx[5] 	& \qw 							& \qw 									& \qw 	\\
\lstick{\ket{q_{ch}}} 		& \qw 	& \ghost{Init} 			& \qw 	& \qw 									& \gate{\delta} \qwx[2] 			& \qw 					& \qw 					& \gate{Reset} \qwx[2] 			& \qw 									& \qw 	\\
\lstick{\ket{q_{tape}}} 	& \qw 	& \ghost{Init} 			& \qw 	& \gate{Read} 							& \qw 								& \multigate{1}{Write} 	& \qw 					& \qw 							& \qw 									& \qw 	\\
\lstick{\ket{q_{write}}} 	& \qw 	& \ghost{Init} 			& \qw 	& \qw 									& \multigate{1}{\delta} 			& \ghost{Write} 		& \qw 					& \multigate{1}{Reset} 			& \qw 									& \qw 	\\
\lstick{\ket{q_{read}}} 	& \qw 	& \ghost{Init} 			& \qw 	& \multigate{1}{Read} \qwx[-2] 			& \ghost{\delta} 					& \qw 					& \qw 					& \ghost{Reset} 				& \qw 									& \qw 	\\
\lstick{\ket{q_{head}}} 	& \qw 	& \ghost{Init} 			& \qw 	& \ghost{Read} 							& \qw 								& \gate{Write} \qwx[-2]	& \multigate{1}{Move} 	& \qw 							& \qw 									& \qw 	\\
\lstick{\ket{q_{move}}} 	& \qw 	& \ghost{Init} 			& \qw 	& \qw 									& \multigate{2}{\delta} \qwx[-2] 	& \qw 					& \ghost{Move} 			& \multigate{2}{Reset} \qwx[-2]	& \qw 									& \qw 	\\
\lstick{\ket{q_{state}}} 	& \qw 	& \ghost{Init} 			& \qw 	& \qw 									& \ghost{\delta} 					& \qw 					& \qw 					& \ghost{Reset} 				& \qw 									& \qw 	\\
\lstick{\ket{q_{\delta}}} 	& \qw 	& \ghost{Init} 			& \qw 	& \qw 									& \ghost{\delta} 					& \qw 					& \qw 					& \ghost{Reset} 				& \qw \gategroup{1}{5}{9}{9}{2em}{.} 	& \qw 	\\
\\
& & & & & & \mbox{QPULBA step} \\
}
}
\caption{Blocks for the quantum circuit implementation of a QPULBA step}
\label{qcirc:qutm}
\end{figure}
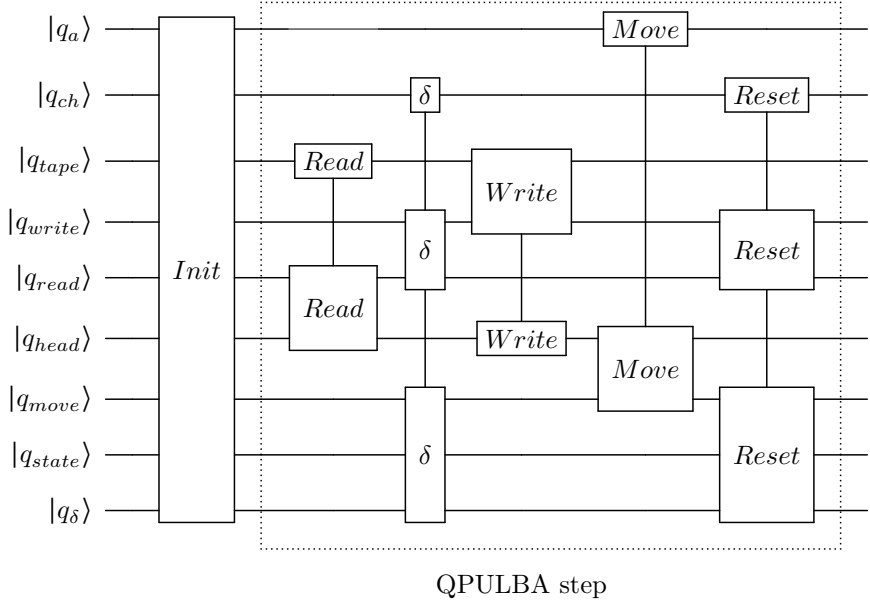

In this section, we present the detailed design of the quantum circuit to implement the ULBA computation model of the previous section.
This is a mechanistic model of a QPULBA having the corresponding parts of a classical ULBA as qubits.
The acronym expansion of `quantum', `parallel', `universal', `linear bounded' translates respectively to the automata features of a superposition in inputs, a superposition of programs, a stored-program model and a memory restricted implementation.
As highlighted in Table~\ref{fig:TMs}, QPULBA (in yellow) captures the computing capabilities of $27$ (in blue) out of $51$ automata models (of type 3, 2, 1) that is realistically implementable on physical hardware.
The circuit in figure~\ref{qcirc:qutm} requires some ancilla qubits which will be introduced later.
% The test qubits for experimenting on the implementation are not part of the logic and thus not shown in figure~\ref{qcirc:qutm}.
The automata step needs to be repeated for the number of steps $t$ we intend to execute the machine before measuring out the qubits.

% \gategroup{t}{l}{b}{r}{.7em}{}

% \begin{figure}[h]
% \centering 
% \centerline{
% }
% \caption{\noteA{add caption and label}}
% %\label{qcirc:}
% \end{figure}

% \newpage
\subsection{Qubit complexity} \label{s6s1}

The qubit complexities of the design elements are discussed here.
The generic formula is derived before applying to the specific case of the 2-2-1 QPULBA of \S~4 \ref{s5s4s6}{}.

\begin{itemize}[nolistsep,noitemsep]
    
    \item \textbf{Alphabet}:
    Alphabet set cardinality $n=|\Gamma|$ is the number of symbols in the alphabet.
    The number of bits/qubits required to represent a symbol, $q_\Gamma = ceil(log_2(n))$
    
    \item \textbf{Head position}:
    The current head position is represented either as binary or one-hot encoding.
    The one-hot encoding is more expensive in the number of qubits, but better in terms of gates.
    The number of bits/qubits required for one-hot encoding~\cite{molina2019revisiting} is the same as the number of cells $c$.
    Since the simulation bottleneck is the number of qubits instead of the number of gates, we prefer the binary encoding.
    For binary encoding, $q_{head} = ceil(log_2(c))$
    
    \item \textbf{Read head}:
    The read head temporarily stores the content of the current head position, requiring bits/qubits, $q_{read} = q_\Gamma$
    
    \item \textbf{Write head}:
    The write head temporarily stores the content to be written to the current head position, requiring bits/qubits, $q_{write} = q_\Gamma$
    
    \item \textbf{Turing tape}:
    The number of bits/qubits required for the restricted tape size of $c$ is, $q_{tape} = c * q_\Gamma $
    
    \item \textbf{Movement}:
    Specifying the movement of a $d$ dimensional Turing tape requires, $q_{move} = d$
    
    \item \textbf{Current state}:
    The current state in binary encoding requires, $q_{state} = ceil(log_2(m))$.
    In a one-hot coded format, it would require $m$ qubits.
    
    \item \textbf{Transition table}:
    The transition function is a unitary matrix that transforms the input and the current state to the output, next state and movement of the tape.
    Thus, for each combination of state and read character, we need to store the next state, write character and movement.
    The number of qubits required are, $q_\delta = (m*n)*(q_{state}+q_{write}+q_{move})$
    
    \item \textbf{Computation history}:
    Since the quantum circuit is reversible, the computation history for $(t-1)$ steps needs to be stored in ancilla qubits.
    The computation history is specified by the state and read symbol for each step, requiring, $q_{ch} = (t-1)*(q_{state}+q_{read})$.
    However, it is possible to trade-off space (qubits) with time (operations) by uncomputing, as discussed in \S~5 \ref{s6s3s5}. 
    
\end{itemize}

% \textbf{Flags}

% These flags are for analytics and are not implemented in the basic version.
% $$q_{flag} = q_{halt}+q_{err}$$

% \subsubsection{Halt indicator}
% This flag indicates if the program has stabilized in an accepting state.
% $$q_{halt} = 1$$

% \subsubsection{Error indicator}
% This flag indicates tape under/overflow. It is however not required if $c \ge (2z+1)$.
% $$q_{err} = 1$$

Thus, the qubit complexity of the implementation (assuming $q_a$ ancilla qubits) is:
\begin{align*} 
q_{QPULBA} &=  q_\delta+q_{state}+q_{move}+q_{head}+q_{read}+q_{write}+q_{tape}+q_{ch}+q_a \\[-0.3em]
        &= (m*n*(log(m)+log(n)+d))+log(m)+d+log(c)+log(n)+log(n)+(c*log(n))\\[-0.5em] & +q_{ch}+q_a
\end{align*}

Considering the 2-2-1 QPULBA example, the values of $m=2$, $n=2$, $d=1$, $c=12$, $t=12$ are substituted in the above equation (all logarithms are base-2 and rounded up to be nearest integer) to yield,

\begin{align*} 
q_{QPULBA}^{221} &= (2*2*(log(2)+log(2)+1))+log(2)+1+log(12)+log(2)+log(2)+(12*log(2))\\[-0.5em] & +(11*(log(2)+log(2)))+q_a\\[-0.3em]
    &= 12+1+1+4+1+1+12+22+q_a\\[-0.3em]
    &= 54+q_a
\end{align*}
Simulating in order of $50$ qubits is near the quantum supremacy limits. 
However, the circuit is not always in full superposition thereby allowing smart simulation techniques in a quantum simulator and uncomputing away the computation history.

% \newpage
\subsection{Initialize} \label{s6s2}

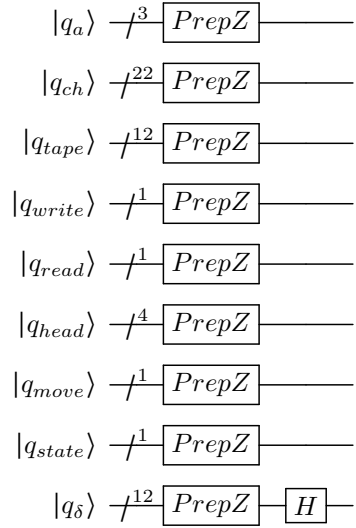
\begin{figure}[htb]
\centering 
\centerline{
\Qcircuit @C=1em @R=0.8em {
\lstick{\ket{q_{a}}}	 	& {/^{3}} 	\qw 	& \gate{PrepZ} & \qw 		& \qw \\
\lstick{\ket{q_{ch}}} 		& {/^{22}} 	\qw 	& \gate{PrepZ} & \qw 		& \qw \\
\lstick{\ket{q_{tape}}} 	& {/^{12}} 	\qw 	& \gate{PrepZ} & \qw 		& \qw \\
\lstick{\ket{q_{write}}} 	& {/^1} 	\qw 	& \gate{PrepZ} & \qw 		& \qw \\
\lstick{\ket{q_{read}}} 	& {/^1} 	\qw 	& \gate{PrepZ} & \qw 		& \qw \\
\lstick{\ket{q_{head}}} 	& {/^4} 	\qw 	& \gate{PrepZ} & \qw 		& \qw \\
\lstick{\ket{q_{move}}} 	& {/^1} 	\qw 	& \gate{PrepZ} & \qw 		& \qw \\
\lstick{\ket{q_{state}}} 	& {/^1} 	\qw 	& \gate{PrepZ} & \qw 		& \qw \\
\lstick{\ket{q_{\delta}}} 	& {/^{12}} 	\qw 	& \gate{PrepZ} & \gate{H} 	& \qw 
}
}
\caption{Initialization quantum circuit for QPULBA 2-2-1}
\label{qcirc:TM221init}
\end{figure}

The initialization circuit depends on the target application for this framework.
For measuring the algorithmic probability or the universal distribution, all possible programs (represented by the transition table) need to be evolved in a superposition.
All other qubits are kept at the ground or default state of $\ket{0}$.
The circuit is shown in figure~\ref{qcirc:TM221init}.

% \newpage
\subsection{QPULBA step} \label{s6s3}

Each iteration of the QPULBA undergoes the following transforms:
\begin{enumerate}[nolistsep,noitemsep]
    \item Read: $\{q_{read}\} \leftarrow U_{read} (\{q_{head}, q_{tape}\})$
	\item Transition evaluation: $\{q_{write}, q_{ch}, q_{move}\} \leftarrow U_{\delta} (\{q_{read}, q_{state}, q_{\delta}\})$
	\item Write: $\{q_{tape}\} \leftarrow U_{write} (\{q_{head}, q_{write}\})$
	\item Move: $\{q_{head}\} \leftarrow U_{move} (\{q_{head}, q_{move}\})$
	\item Reset
\end{enumerate}

This corresponds to one step of a classical UTM, with the distinction of the computation now evolving in a superposition of all possible classical automata. We will now discuss each of these QPULBA steps in detail.

\newpage
\subsubsection{Read tape} \label{s6s3s1}

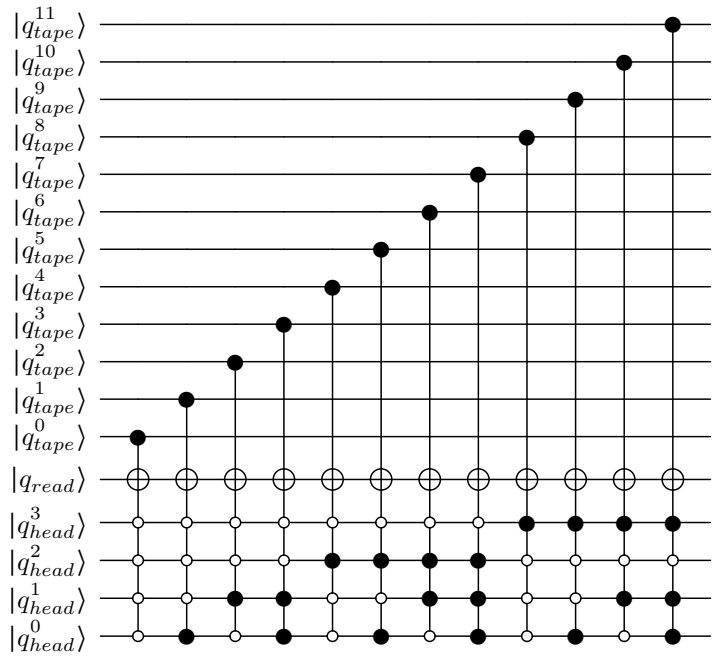
\begin{figure}[htb]
\centering 
\centerline{
\Qcircuit @C=1.0em @R=1.0em {
\lstick{\ket{q_{tape}^{11}}}    & \qw           & \qw           & \qw           & \qw           & \qw           & \qw           & \qw           & \qw           & \qw           & \qw           & \qw           & \ctrl{12}     & \qw \\
\lstick{\ket{q_{tape}^{10}}}    & \qw           & \qw           & \qw           & \qw           & \qw           & \qw           & \qw           & \qw           & \qw           & \qw           & \ctrl{11}     & \qw           & \qw \\
\lstick{\ket{q_{tape}^9}}       & \qw           & \qw           & \qw           & \qw           & \qw           & \qw           & \qw           & \qw           & \qw           & \ctrl{10}     & \qw           & \qw           & \qw \\
\lstick{\ket{q_{tape}^8}}       & \qw           & \qw           & \qw           & \qw           & \qw           & \qw           & \qw           & \qw           & \ctrl{9}      & \qw           & \qw           & \qw           & \qw \\
\lstick{\ket{q_{tape}^7}}       & \qw           & \qw           & \qw           & \qw           & \qw           & \qw           & \qw           & \ctrl{8}      & \qw           & \qw           & \qw           & \qw	        & \qw \\
\lstick{\ket{q_{tape}^6}}       & \qw           & \qw           & \qw           & \qw           & \qw           & \qw           & \ctrl{7}      & \qw           & \qw           & \qw           & \qw           & \qw           & \qw \\
\lstick{\ket{q_{tape}^5}}       & \qw           & \qw           & \qw           & \qw           & \qw           & \ctrl{6}      & \qw           & \qw           & \qw           & \qw           & \qw           & \qw           & \qw \\
\lstick{\ket{q_{tape}^4}}       & \qw           & \qw           & \qw           & \qw           & \ctrl{5}      & \qw           & \qw           & \qw           & \qw           & \qw           & \qw           & \qw           & \qw \\
\lstick{\ket{q_{tape}^3}}       & \qw           & \qw           & \qw           & \ctrl{4}      & \qw           & \qw           & \qw           & \qw           & \qw           & \qw           & \qw           & \qw           & \qw \\
\lstick{\ket{q_{tape}^2}}       & \qw           & \qw           & \ctrl{3}      & \qw           & \qw           & \qw           & \qw           & \qw           & \qw           & \qw           & \qw           & \qw           & \qw \\
\lstick{\ket{q_{tape}^1}}       & \qw           & \ctrl{2}      & \qw           & \qw           & \qw           & \qw           & \qw           & \qw           & \qw           & \qw           & \qw           & \qw           & \qw \\
\lstick{\ket{q_{tape}^0}}       & \ctrl{1}      & \qw           & \qw           & \qw           & \qw           & \qw           & \qw           & \qw           & \qw           & \qw           & \qw           & \qw           & \qw \\
\lstick{\ket{q_{read}}}         & \targ         & \targ         & \targ         & \targ         & \targ         & \targ         & \targ         & \targ         & \targ         & \targ         & \targ         & \targ         & \qw \\
\lstick{\ket{q_{head}^3}}       & \ctrlo{-1}    & \ctrlo{-1}    & \ctrlo{-1}    & \ctrlo{-1}    & \ctrlo{-1}    & \ctrlo{-1}    & \ctrlo{-1}    & \ctrlo{-1}    & \ctrl{-1}     & \ctrl{-1}     & \ctrl{-1}     & \ctrl{-1}     & \qw \\
\lstick{\ket{q_{head}^2}}       & \ctrlo{-1}    & \ctrlo{-1}    & \ctrlo{-1}    & \ctrlo{-1}    & \ctrl{-1}     & \ctrl{-1}     & \ctrl{-1}     & \ctrl{-1}     & \ctrlo{-1}    & \ctrlo{-1}    & \ctrlo{-1}    & \ctrlo{-1}    & \qw \\
\lstick{\ket{q_{head}^1}}       & \ctrlo{-1}    & \ctrlo{-1}    & \ctrl{-1}     & \ctrl{-1}     & \ctrlo{-1}    & \ctrlo{-1}    & \ctrl{-1}     & \ctrl{-1}     & \ctrlo{-1}    & \ctrlo{-1}    & \ctrl{-1}     & \ctrl{-1}     & \qw \\
\lstick{\ket{q_{head}^0}}       & \ctrlo{-1}    & \ctrl{-1}     & \ctrlo{-1}    & \ctrl{-1}     & \ctrlo{-1}    & \ctrl{-1}     & \ctrlo{-1}    & \ctrl{-1}     & \ctrlo{-1}    & \ctrl{-1}     & \ctrlo{-1}    & \ctrl{-1}     & \qw \\
}
}
\caption{Read tape quantum circuit for QPULBA 2-2-1}
\label{qcirc:read}
\end{figure}

The quantum circuit implements a multiplexer with the tape as the input signals and the binary coded head position as the selector lines. 
The read head is the output.
% $$\Qcircuit @C=1.0em @R=1.0em {
% \lstick{\ket{q_{tape}}} & {/} \qw   &   \ctrlm{1}  & \qw       \\
% \lstick{\ket{q_{read}}} & \qw       &   \targ   & \qw       \\
% \lstick{\ket{q_{head}}} & {/} \qw       &   \ctrle{-1}  & \qw 
% }$$
The read circuit for the QPULBA 2-2-1 is shown in figure~\ref{qcirc:read}.

\newpage
\subsubsection{Transition table lookup} \label{s6s3s2}

The transition table encoding is: $[Q_{t}|R_\Gamma] \rightarrow [Q_{t+1}|M_{l/r}|W_\Gamma]$

% $$\Qcircuit @C=1.0em @R=1.0em {
% \lstick{\ket{q_{state}^{-}}} & \qw   &   \targm  & \qw       \\
% \lstick{\ket{q_{write}}} & \qw       &   \targm   & \qw       \\
% \lstick{\ket{q_{read}}} & \qw       &   \ctrlm{-2}  & \qw       \\
% \lstick{\ket{q_{move}}} & \qw       &   \targm  & \qw       \\
% \lstick{\ket{q_{state}}} & \qw       &   \ctrlb{-2}  & \qw       \\
% \lstick{\ket{q_{\delta}}} & {/} \qw       &   \ctrlm{-1}  & \qw       \\
% }$$

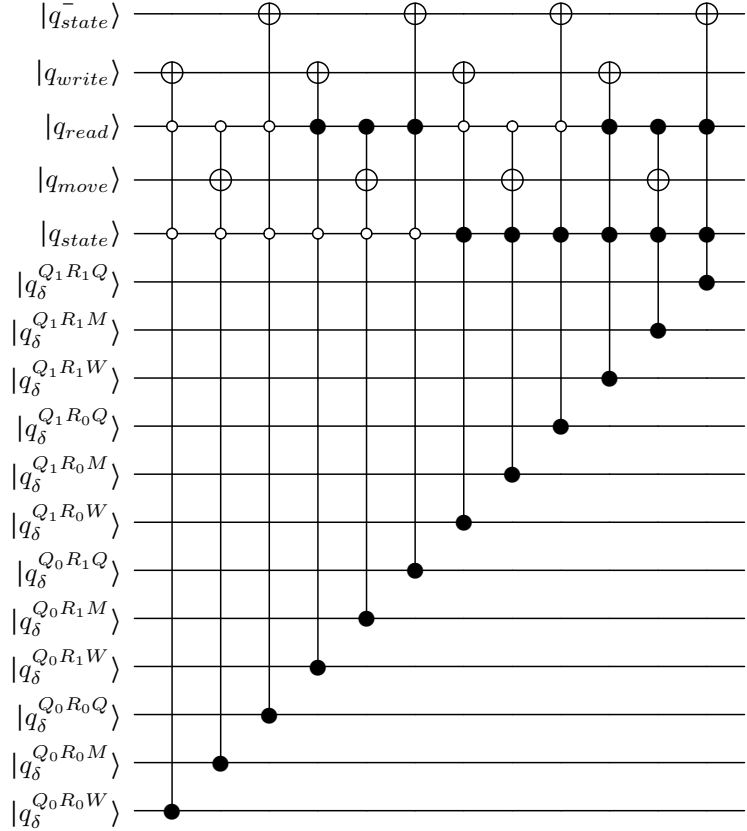
\begin{figure}[htb]
\centering 
\centerline{
\Qcircuit @C=1.0em @R=1.4em {
\lstick{\ket{q_{state}^{-}}} 		&  \qw         & \qw         & \targ       & \qw         & \qw         & \targ       & \qw         & \qw         & \targ       & \qw         & \qw         & \targ       & \qw \\
\lstick{\ket{q_{write}}} 			&  \targ       & \qw         & \qw         & \targ       & \qw         & \qw         & \targ       & \qw         & \qw         & \targ       & \qw         & \qw         & \qw \\
\lstick{\ket{q_{read}}} 			&  \ctrlo{-1}  & \ctrlo{1}   & \ctrlo{-2}  & \ctrl{-1}   & \ctrl{1}    & \ctrl{-2}   & \ctrlo{-1}  & \ctrlo{1}   & \ctrlo{-2}  & \ctrl{-1}   & \ctrl{1}    & \ctrl{-2}   & \qw \\
\lstick{\ket{q_{move}}} 			&  \qw         & \targ       & \qw         & \qw         & \targ       & \qw         & \qw         & \targ       & \qw         & \qw         & \targ       & \qw         & \qw \\
\lstick{\ket{q_{state}}} 			&  \ctrlo{-2}  & \ctrlo{-1}  & \ctrlo{-2}  & \ctrlo{-2}  & \ctrlo{-1}  & \ctrlo{-2}  & \ctrl{-2}   & \ctrl{-1}   & \ctrl{-2}   & \ctrl{-2}   & \ctrl{-1}   & \ctrl{-2}   & \qw \\
\lstick{\ket{q_{\delta}^{Q_1R_1Q}}} &  \qw         & \qw         & \qw         & \qw         & \qw         & \qw         & \qw         & \qw         & \qw         & \qw         & \qw         & \ctrl{-1}   & \qw \\
\lstick{\ket{q_{\delta}^{Q_1R_1M}}} &  \qw         & \qw         & \qw         & \qw         & \qw         & \qw         & \qw         & \qw         & \qw         & \qw         & \ctrl{-2}   & \qw         & \qw \\
\lstick{\ket{q_{\delta}^{Q_1R_1W}}} &  \qw         & \qw         & \qw         & \qw         & \qw         & \qw         & \qw         & \qw         & \qw         & \ctrl{-3}   & \qw         & \qw         & \qw \\
\lstick{\ket{q_{\delta}^{Q_1R_0Q}}} &  \qw         & \qw         & \qw         & \qw         & \qw         & \qw         & \qw         & \qw         & \ctrl{-4}   & \qw         & \qw         & \qw         & \qw \\
\lstick{\ket{q_{\delta}^{Q_1R_0M}}} &  \qw         & \qw         & \qw         & \qw         & \qw         & \qw         & \qw         & \ctrl{-5}   & \qw         & \qw         & \qw         & \qw         & \qw \\
\lstick{\ket{q_{\delta}^{Q_1R_0W}}} &  \qw         & \qw         & \qw         & \qw         & \qw         & \qw         & \ctrl{-6}   & \qw         & \qw         & \qw         & \qw         & \qw         & \qw \\
\lstick{\ket{q_{\delta}^{Q_0R_1Q}}} &  \qw         & \qw         & \qw         & \qw         & \qw         & \ctrl{-7}   & \qw         & \qw         & \qw         & \qw         & \qw         & \qw         & \qw \\
\lstick{\ket{q_{\delta}^{Q_0R_1M}}} &  \qw         & \qw         & \qw         & \qw         & \ctrl{-8}   & \qw         & \qw         & \qw         & \qw         & \qw         & \qw         & \qw         & \qw \\
\lstick{\ket{q_{\delta}^{Q_0R_1W}}} &  \qw         & \qw         & \qw         & \ctrl{-9}   & \qw         & \qw         & \qw         & \qw         & \qw         & \qw         & \qw         & \qw         & \qw \\
\lstick{\ket{q_{\delta}^{Q_0R_0Q}}} &  \qw         & \qw         & \ctrl{-10}  & \qw         & \qw         & \qw         & \qw         & \qw         & \qw         & \qw         & \qw         & \qw         & \qw \\
\lstick{\ket{q_{\delta}^{Q_0R_0M}}} &  \qw         & \ctrl{-11}  & \qw         & \qw         & \qw         & \qw         & \qw         & \qw         & \qw         & \qw         & \qw         & \qw         & \qw \\
\lstick{\ket{q_{\delta}^{Q_0R_0W}}} &  \ctrl{-12}  & \qw         & \qw         & \qw         & \qw         & \qw         & \qw         & \qw         & \qw         & \qw         & \qw         & \qw         & \qw \\
}
}
\caption{Transition function quantum circuit for QPULBA 2-2-1}
\label{qcirc:fsm}
\end{figure}

Note that we use $q_{state}^{-}$ instead of $q_{state}^{+}$ though we are storing the next state in the qubit.
This is corrected by the reset circuit.
The transition function circuit for the QPULBA 2-2-1 is shown in figure~\ref{qcirc:fsm}.

\newpage
\subsubsection{Write tape} \label{s6s3s3}

The quantum circuit implements a de-multiplexer with the tape as the output signals and the head position as the selector lines. 
The write head is the input.

% $$\Qcircuit @C=1.0em @R=1.0em {
% \lstick{\ket{q_{tape}}} & {/} \qw   &   \targm  & \qw       \\
% \lstick{\ket{q_{write}}} & \qw       &   \ctrl{-1}   & \qw       \\
% \lstick{\ket{q_{head}}} & {/} \qw       &   \ctrlb{-1}  & \qw       \\
% }$$

\begin{figure}[htb]
\centering 
\centerline{
\Qcircuit @C=1.0em @R=1.0em {
\lstick{\ket{q_{tape}^{11}}}    & \qw           & \qw           & \qw           & \qw           & \qw           & \qw           & \qw           & \qw           & \qw           & \qw           & \qw           & \targ         & \qw \\
\lstick{\ket{q_{tape}^{10}}}    & \qw           & \qw           & \qw           & \qw           & \qw           & \qw           & \qw           & \qw           & \qw           & \qw           & \targ         & \qw           & \qw \\
\lstick{\ket{q_{tape}^9}}       & \qw           & \qw           & \qw           & \qw           & \qw           & \qw           & \qw           & \qw           & \qw           & \targ         & \qw           & \qw           & \qw \\
\lstick{\ket{q_{tape}^8}}       & \qw           & \qw           & \qw           & \qw           & \qw           & \qw           & \qw           & \qw           & \targ         & \qw           & \qw           & \qw           & \qw \\
\lstick{\ket{q_{tape}^7}}       & \qw           & \qw           & \qw           & \qw           & \qw           & \qw           & \qw           & \targ         & \qw           & \qw           & \qw           & \qw	        & \qw \\
\lstick{\ket{q_{tape}^6}}       & \qw           & \qw           & \qw           & \qw           & \qw           & \qw           & \targ         & \qw           & \qw           & \qw           & \qw           & \qw           & \qw \\
\lstick{\ket{q_{tape}^5}}       & \qw           & \qw           & \qw           & \qw           & \qw           & \targ         & \qw           & \qw           & \qw           & \qw           & \qw           & \qw           & \qw \\
\lstick{\ket{q_{tape}^4}}       & \qw           & \qw           & \qw           & \qw           & \targ         & \qw           & \qw           & \qw           & \qw           & \qw           & \qw           & \qw           & \qw \\
\lstick{\ket{q_{tape}^3}}       & \qw           & \qw           & \qw           & \targ         & \qw           & \qw           & \qw           & \qw           & \qw           & \qw           & \qw           & \qw           & \qw \\
\lstick{\ket{q_{tape}^2}}       & \qw           & \qw           & \targ         & \qw           & \qw           & \qw           & \qw           & \qw           & \qw           & \qw           & \qw           & \qw           & \qw \\
\lstick{\ket{q_{tape}^1}}       & \qw           & \targ         & \qw           & \qw           & \qw           & \qw           & \qw           & \qw           & \qw           & \qw           & \qw           & \qw           & \qw \\
\lstick{\ket{q_{tape}^0}}       & \targ         & \qw           & \qw           & \qw           & \qw           & \qw           & \qw           & \qw           & \qw           & \qw           & \qw           & \qw           & \qw \\
\lstick{\ket{q_{write}}}         & \ctrl{-1}     & \ctrl{-2}     & \ctrl{-3}     & \ctrl{-4}     & \ctrl{-5}     & \ctrl{-6}     & \ctrl{-7}     & \ctrl{-8}     & \ctrl{-9}     & \ctrl{-10}    & \ctrl{-11}    & \ctrl{-12}    & \qw \\
\lstick{\ket{q_{head}^3}}       & \ctrlo{-1}    & \ctrlo{-1}    & \ctrlo{-1}    & \ctrlo{-1}    & \ctrlo{-1}    & \ctrlo{-1}    & \ctrlo{-1}    & \ctrlo{-1}    & \ctrl{-1}     & \ctrl{-1}     & \ctrl{-1}     & \ctrl{-1}     & \qw \\
\lstick{\ket{q_{head}^2}}       & \ctrlo{-1}    & \ctrlo{-1}    & \ctrlo{-1}    & \ctrlo{-1}    & \ctrl{-1}     & \ctrl{-1}     & \ctrl{-1}     & \ctrl{-1}     & \ctrlo{-1}    & \ctrlo{-1}    & \ctrlo{-1}    & \ctrlo{-1}    & \qw \\
\lstick{\ket{q_{head}^1}}       & \ctrlo{-1}    & \ctrlo{-1}    & \ctrl{-1}     & \ctrl{-1}     & \ctrlo{-1}    & \ctrlo{-1}    & \ctrl{-1}     & \ctrl{-1}     & \ctrlo{-1}    & \ctrlo{-1}    & \ctrl{-1}     & \ctrl{-1}     & \qw \\
\lstick{\ket{q_{head}^0}}       & \ctrlo{-1}    & \ctrl{-1}     & \ctrlo{-1}    & \ctrl{-1}     & \ctrlo{-1}    & \ctrl{-1}     & \ctrlo{-1}    & \ctrl{-1}     & \ctrlo{-1}    & \ctrl{-1}     & \ctrlo{-1}    & \ctrl{-1}     & \qw \\
}
}
\caption{Write tape quantum circuit for QPULBA 2-2-1}
\label{qcirc:write}
\end{figure}
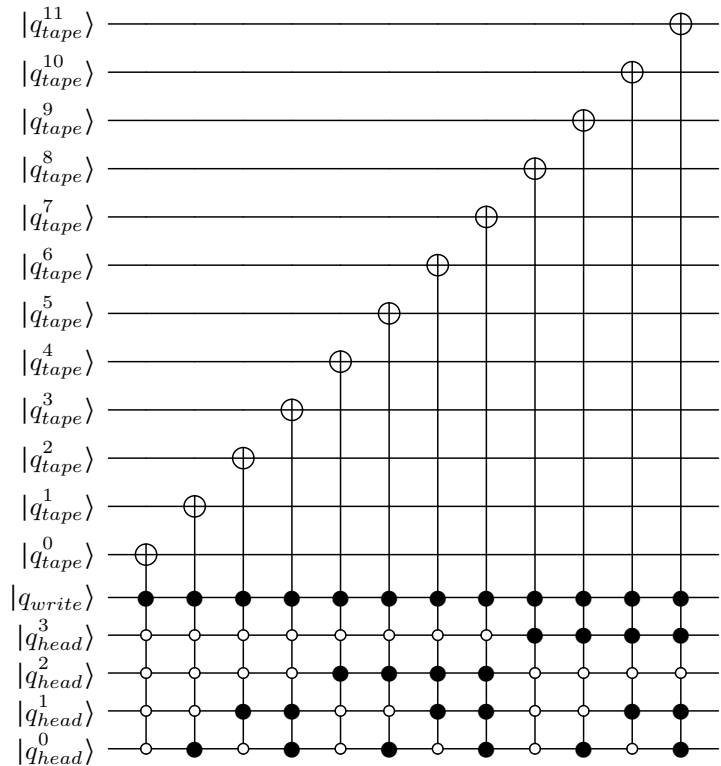

The write circuit for the QPULBA 2-2-1 is shown in figure~\ref{qcirc:write}.
The qubit elements not involved are not shown.

\newpage
\subsubsection{Move} \label{s6s3s4}

There are many choices for implementing the move, e.g. a looped tape (overflow/underflow is ignored and trimmed), error flag is raised and halts, overflow/underflow is ignored, etc.
Here, a looped tape is implemented.
    
The head is incremented/decremented using the move qubit as control.
$$\Qcircuit @C=1.0em @R=1.0em {
\lstick{\ket{q_{head}}} & {/} \qw   &   \gate{Inc}  & \qw       &   \gate{Dec}  & \qw       & \qw       \\
\lstick{\ket{q_{move}}} & \qw       &   \ctrl{-1}   & \gate{X}  &   \ctrl{-1}   & \gate{X}  & \qw       \\
}$$

The increment/decrement circuit is a special case of the quantum full adder~\cite{vedral1996quantum} with the first register, $a$ set to $1$.
For the QPULBA 2-2-1 case, the length of the circular tape is $12$, thus, the increment and decrement needs to be modulo $12$.
For increment, when the $q_{head}$ equals $11$, it should increment to $(11+1) \text{ mod } 12 = 0$, while for decrement, $(0-1)\text{ mod }12 = 11$.
Thus, for these edge cases, we need to increment/decrement by $5$ instead of $1$.
We set the $a_2$ bit to change the effective value of $a$ from $1=0001_2$ to $5=0101_2$ for the addition/subtraction.
This operation is conditioned on the head value and move bit, and denoted as the overflow/underflow qubit $\ket{ovfw}$/$\ket{udfw}$.
The $a_2$ bit is uncomputed based on the incremented value being $0$ or the decremented value being $11$.

The carry ($C$), sum ($S$) and reverse carry ($C^\dagger$) blocks are defined as follows:
\begin{verbatim}
        .sum                .carry                  .reverse_carry
        cnot A0,S0          toffoli A0,B0,C1        toffoli C0,B0,C1
        cnot B0,S0          cnot A0,B0              cnot A0,B0
                            toffoli C0,B0,C1        toffoli A0,B0,C1
\end{verbatim}

\newpage
\textbf{Increment}

\begin{figure}[htb]
\centering 
\centerline{
\Qcircuit @C=0.6em @R=0.6em {
\lstick{\ket{0}=c_0} 			& \qw				& \multigate{3}{C}	& \qw       		& \qw      			& \qw      		 	& \qw      		 	& \qw      		 	& \qw      		 			& \qw      		 	& \qw      		 			& \qw      		 	& \multigate{3}{C^\dagger}	& \multigate{2}{S}	& \qw				& \qw      		 	& \rstick{\ket{0}}			\\
\lstick{\ket{q_{move}}=a_0} 	& \ctrl{1}			& \ghost{C}			& \qw       		& \qw      			& \qw      		 	& \qw      		 	& \qw      		 	& \qw      		 			& \qw      		 	& \qw      		 			& \qw      		 	& \ghost{C^\dagger}		  	& \ghost{S}		 	& \ctrl{1}			& \qw      		 	& \rstick{\ket{q_{move}}}	\\
\lstick{\ket{q_{head}^0}=b_0} 	& \ctrl{3}			& \ghost{C}			& \qw       		& \qw      			& \qw      		 	& \qw      		 	& \qw      		 	& \qw      		 			& \qw      		 	& \qw      		 			& \qw      		 	& \ghost{C^\dagger}		  	& \ghost{S}		  	& \ctrlo{3}			& \qw      		 	\\
\lstick{\ket{0}=c_1} 			& \qw				& \ghost{C}			& \multigate{3}{C}	& \qw      			& \qw      		 	& \qw      		 	& \qw      		 	& \qw      		 			& \qw      		 	& \multigate{3}{C^\dagger}	& \multigate{2}{S}	& \ghost{C^\dagger}		  	& \qw      		 	& \qw				& \qw      		 	& \rstick{\ket{0}}			\\
\lstick{\ket{0}=a_1} 			& \qw				& \qw				& \ghost{C}		    & \qw      			& \qw      		 	& \qw      		 	& \qw      		 	& \qw      		 			& \qw      		 	& \ghost{C^\dagger}		  	& \ghost{S}		 	& \qw      		 			& \qw      		 	& \qw				& \qw      		 	& \rstick{\ket{0}}			\\
\lstick{\ket{q_{head}^1}=b_1} 	& \ctrl{2}			& \qw				& \ghost{C}		    & \qw      			& \qw      		 	& \qw      		 	& \qw      		 	& \qw      		 			& \qw      		 	& \ghost{C^\dagger}		  	& \ghost{S}		  	& \qw      		 			& \qw      		 	& \ctrlo{2}			& \qw      		 	\\
\lstick{\ket{0}=c_2} 			& \qw				& \qw				& \ghost{C}		    & \multigate{3}{C}	& \qw      		 	& \qw      		 	& \qw      		 	& \multigate{3}{C^\dagger}	& \multigate{2}{S}	& \ghost{C^\dagger}		  	& \qw      		 	& \qw      		 			& \qw      		 	& \qw				& \qw      		 	& \rstick{\ket{0}}			\\
\lstick{\ket{ovfw}=a_2} 		& \targ				& \qw				& \qw			    & \ghost{C}		  	& \qw      		 	& \qw      		 	& \qw      		 	& \ghost{C^\dagger}		  	& \ghost{S}		 	& \qw      		 			& \qw      		 	& \qw      		 			& \qw      		 	& \targ				& \qw      		 	& \rstick{\ket{0}}			\\
\lstick{\ket{q_{head}^2}=b_2} 	& \ctrlo{-1}		& \qw				& \qw			    & \ghost{C}		  	& \qw      		 	& \qw      		 	& \qw      		 	& \ghost{C^\dagger}		  	& \ghost{S}		  	& \qw      		 			& \qw      		 	& \qw      		 			& \qw      		 	& \ctrlo{-1}		& \qw      		 	\\
\lstick{\ket{0}=c_3} 			& \qw				& \qw				& \qw			    & \ghost{C}		  	& \multigate{3}{C}	& \qw      		 	& \multigate{2}{S}	& \ghost{C^\dagger}		  	& \qw      		 	& \qw      		 			& \qw      		 	& \qw      		 			& \qw      		 	& \qw				& \qw      		 	& \rstick{\ket{0}}			\\
\lstick{\ket{0}=a_3} 			& \qw				& \qw				& \qw			    & \qw			   	& \ghost{C}		  	& \ctrl{1} 		 	& \ghost{S}		 	& \qw      		 			& \qw      		 	& \qw      		 			& \qw      		 	& \qw      		 			& \qw      		 	& \qw				& \qw      		 	& \rstick{\ket{0}}			\\
\lstick{\ket{q_{head}^3}=b_3} 	& \ctrl{-3}			& \qw				& \qw			    & \qw			   	& \ghost{C}		  	& \targ    		 	& \ghost{S}		  	& \qw      		 			& \qw      		 	& \qw      		 			& \qw      		 	& \qw      		 			& \qw      		 	& \ctrlo{-3}			& \qw      		 	\\
\lstick{\ket{0}=b_4} 			& \qw				& \qw				& \qw			    & \qw			   	& \ghost{C}		  	& \qw      		 	& \qw      		 	& \qw      		 			& \qw      		 	& \qw      		 			& \qw      		 	& \qw      		 			& \qw      		 	& \qw				& \qw      		 	\\
}
}
\caption{Modulo-12 quantum adder for implementing move tape head for QPULBA 2-2-1}
\label{qcirc:addr}
\end{figure}
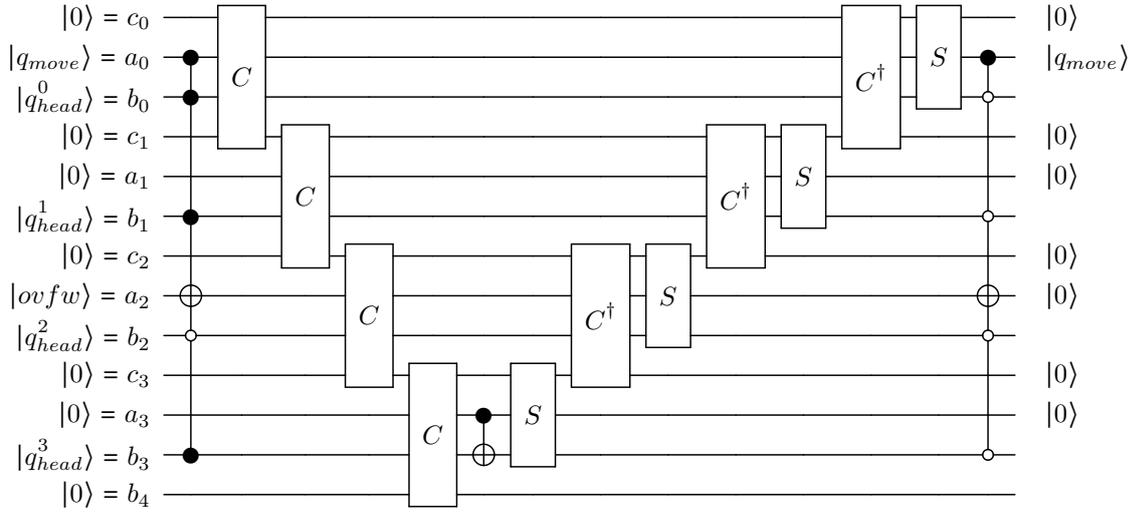

In this design, $c_3c_2c_1 = 000$ are 3 ancilla ($c_0$ is not required), $a_0 = q_{move}$, $a_3a_2a_1 = 000$, $b_3b_2b_1b_0 = q_{head}^3q_{head}^2q_{head}^1q_{head}^0$ and $b_4$ is ignored.
The circuit in figure~\ref{qcirc:addr} shows the 4-bit modulo-12 quantum increment circuit using the quantum adder blocks.

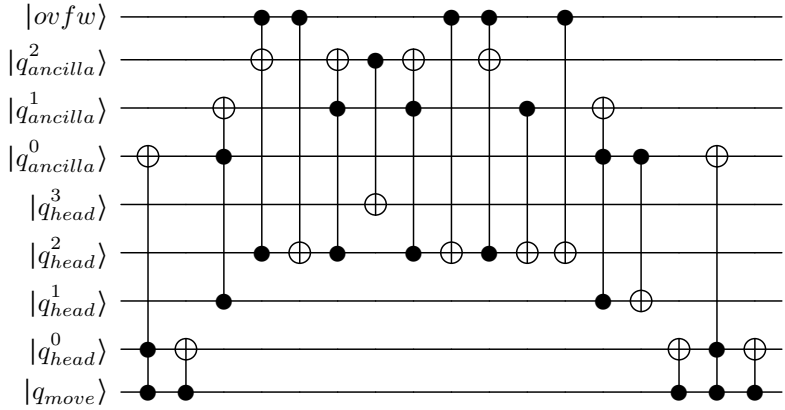
\begin{figure}[htb]
\centering 
\centerline{
\Qcircuit @C=0.6em @R=1.0em {
\lstick{\ket{ovfw}}    			& \qw       & \qw       & \qw       & \ctrl{1}  & \ctrl{5}  & \qw       & \qw       & \qw       & \ctrl{5}  & \ctrl{1}  & \qw       & \ctrl{5}  & \qw       & \qw       & \qw       & \qw       & \qw       & \qw \\
\lstick{\ket{q_{ancilla}^2}}    & \qw       & \qw       & \qw       & \targ     & \qw       & \targ     & \ctrl{3}  & \targ     & \qw       & \targ     & \qw       & \qw       & \qw       & \qw       & \qw       & \qw       & \qw       & \qw \\
\lstick{\ket{q_{ancilla}^1}}    & \qw       & \qw       & \targ     & \qw       & \qw       & \ctrl{-1} & \qw       & \ctrl{-1} & \qw       & \qw       & \ctrl{3}  & \qw       & \targ     & \qw       & \qw       & \qw       & \qw       & \qw \\
\lstick{\ket{q_{ancilla}^0}}    & \targ     & \qw       & \ctrl{-1} & \qw       & \qw       & \qw       & \qw       & \qw       & \qw       & \qw       & \qw       & \qw       & \ctrl{-1} & \ctrl{3}  & \qw       & \targ     & \qw       & \qw \\
\lstick{\ket{q_{head}^3}}       & \qw       & \qw       & \qw       & \qw       & \qw       & \qw       & \targ     & \qw       & \qw       & \qw       & \qw       & \qw       & \qw       & \qw       & \qw       & \qw       & \qw       & \qw \\
\lstick{\ket{q_{head}^2}}       & \qw       & \qw       & \qw       & \ctrl{-4} & \targ     & \ctrl{-3} & \qw       & \ctrl{-3} & \targ     & \ctrl{-4} & \targ     & \targ     & \qw       & \qw       & \qw       & \qw       & \qw       & \qw \\
\lstick{\ket{q_{head}^1}}       & \qw       & \qw       & \ctrl{-3} & \qw       & \qw       & \qw       & \qw       & \qw       & \qw       & \qw       & \qw       & \qw       & \ctrl{-3} & \targ     & \qw       & \qw       & \qw       & \qw \\
\lstick{\ket{q_{head}^0}}       & \ctrl{-4} & \targ     & \qw       & \qw       & \qw       & \qw       & \qw       & \qw       & \qw       & \qw       & \qw       & \qw       & \qw       & \qw       & \targ     & \ctrl{-4} & \targ     & \qw \\
\lstick{\ket{q_{move}}}         & \ctrl{-1} & \ctrl{-1} & \qw       & \qw       & \qw       & \qw       & \qw       & \qw       & \qw       & \qw       & \qw       & \qw       & \qw       & \qw       & \ctrl{-1} & \ctrl{-1} & \ctrl{-1} & \qw \\
}
}
\caption{Modulo-12 increment quantum circuit for QPULBA 2-2-1}
\label{qcirc:inc}
\end{figure}

The circuit can be simplified considering some of the control qubits are always in the $0$ state, so those CNOT/Toffoli gates can be ignored.
The optimized increment circuit for the QPULBA 2-2-1 is shown in figure~\ref{qcirc:inc}.

\vfill
\newpage
\textbf{Decrement}

\begin{figure}[htb]
\centering 
\centerline{
\Qcircuit @C=0.6em @R=0.6em {
\lstick{\ket{0}=c_0} 			& \qw				& \multigate{3}{C}	& \qw       		& \qw      			& \qw      		 	& \qw      		 	& \qw      		 	& \qw      		 			& \qw      		 	& \qw      		 			& \qw      		 	& \multigate{3}{C^\dagger}	& \multigate{2}{S}	& \qw				& \qw      		 	& \rstick{\ket{0}}			\\
\lstick{\ket{q_{move}}=a_0} 	& \ctrl{1}			& \ghost{C}			& \qw       		& \qw      			& \qw      		 	& \qw      		 	& \qw      		 	& \qw      		 			& \qw      		 	& \qw      		 			& \qw      		 	& \ghost{C^\dagger}		  	& \ghost{S}		 	& \ctrl{1}			& \qw      		 	& \rstick{\ket{q_{move}}}	\\
\lstick{\ket{q_{head}^0}=b_0} 	& \ctrlo{3}			& \ghost{C}			& \qw       		& \qw      			& \qw      		 	& \qw      		 	& \qw      		 	& \qw      		 			& \qw      		 	& \qw      		 			& \qw      		 	& \ghost{C^\dagger}		  	& \ghost{S}		  	& \ctrl{3}			& \qw      		 	\\
\lstick{\ket{0}=c_1} 			& \qw				& \ghost{C}			& \multigate{3}{C}	& \qw      			& \qw      		 	& \qw      		 	& \qw      		 	& \qw      		 			& \qw      		 	& \multigate{3}{C^\dagger}	& \multigate{2}{S}	& \ghost{C^\dagger}		  	& \qw      		 	& \qw				& \qw      		 	& \rstick{\ket{0}}			\\
\lstick{\ket{0}=a_1} 			& \qw				& \qw				& \ghost{C}		    & \qw      			& \qw      		 	& \qw      		 	& \qw      		 	& \qw      		 			& \qw      		 	& \ghost{C^\dagger}		  	& \ghost{S}		 	& \qw      		 			& \qw      		 	& \qw				& \qw      		 	& \rstick{\ket{0}}			\\
\lstick{\ket{q_{head}^1}=b_1} 	& \ctrlo{2}			& \qw				& \ghost{C}		    & \qw      			& \qw      		 	& \qw      		 	& \qw      		 	& \qw      		 			& \qw      		 	& \ghost{C^\dagger}		  	& \ghost{S}		  	& \qw      		 			& \qw      		 	& \ctrl{2}			& \qw      		 	\\
\lstick{\ket{0}=c_2} 			& \qw				& \qw				& \ghost{C}		    & \multigate{3}{C}	& \qw      		 	& \qw      		 	& \qw      		 	& \multigate{3}{C^\dagger}	& \multigate{2}{S}	& \ghost{C^\dagger}		  	& \qw      		 	& \qw      		 			& \qw      		 	& \qw				& \qw      		 	& \rstick{\ket{0}}			\\
\lstick{\ket{udfw}=a_2} 		& \targ				& \qw				& \qw			    & \ghost{C}		  	& \qw      		 	& \qw      		 	& \qw      		 	& \ghost{C^\dagger}		  	& \ghost{S}		 	& \qw      		 			& \qw      		 	& \qw      		 			& \qw      		 	& \targ				& \qw      		 	& \rstick{\ket{0}}			\\
\lstick{\ket{q_{head}^2}=b_2} 	& \ctrlo{-1}		& \qw				& \qw			    & \ghost{C}		  	& \qw      		 	& \qw      		 	& \qw      		 	& \ghost{C^\dagger}		  	& \ghost{S}		  	& \qw      		 			& \qw      		 	& \qw      		 			& \qw      		 	& \ctrlo{-1}		& \qw      		 	\\
\lstick{\ket{0}=c_3} 			& \qw				& \qw				& \qw			    & \ghost{C}		  	& \multigate{3}{C}	& \qw      		 	& \multigate{2}{S}	& \ghost{C^\dagger}		  	& \qw      		 	& \qw      		 			& \qw      		 	& \qw      		 			& \qw      		 	& \qw				& \qw      		 	& \rstick{\ket{0}}			\\
\lstick{\ket{0}=a_3} 			& \qw				& \qw				& \qw			    & \qw			   	& \ghost{C}		  	& \ctrl{1} 		 	& \ghost{S}		 	& \qw      		 			& \qw      		 	& \qw      		 			& \qw      		 	& \qw      		 			& \qw      		 	& \qw				& \qw      		 	& \rstick{\ket{0}}			\\
\lstick{\ket{q_{head}^3}=b_3} 	& \ctrlo{-3}		& \qw				& \qw			    & \qw			   	& \ghost{C}		  	& \targ    		 	& \ghost{S}		  	& \qw      		 			& \qw      		 	& \qw      		 			& \qw      		 	& \qw      		 			& \qw      		 	& \ctrl{-3}			& \qw      		 	\\
\lstick{\ket{0}=b_4} 			& \qw				& \qw				& \qw			    & \qw			   	& \ghost{C}		  	& \qw      		 	& \qw      		 	& \qw      		 			& \qw      		 	& \qw      		 			& \qw      		 	& \qw      		 			& \qw      		 	& \qw				& \qw      		 	\\
}
}
\caption{Modulo-12 quantum subtractor for implementing move tape head for QPULBA 2-2-1}
\label{qcirc:subt}
\end{figure}
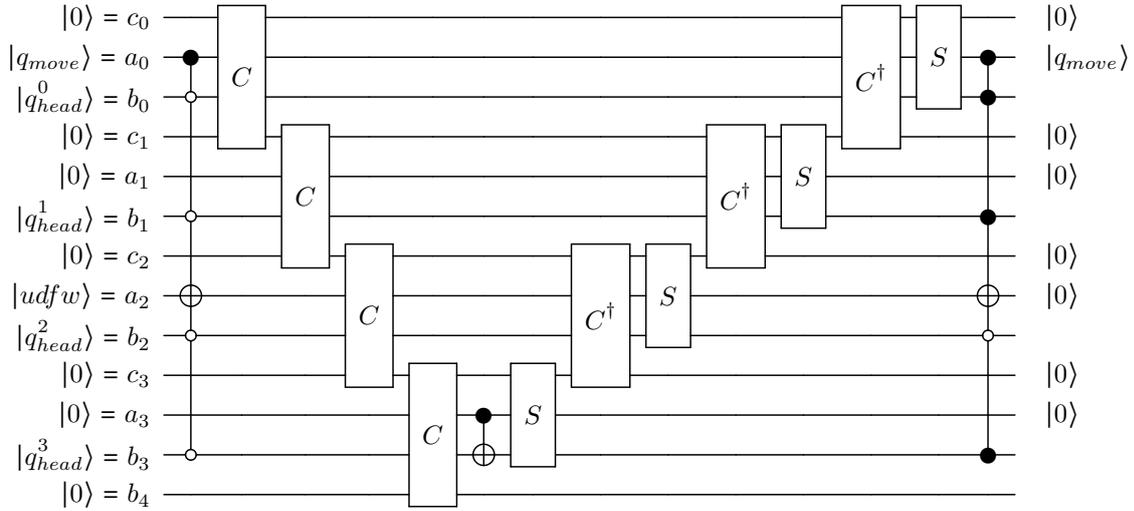

In this design, $c_3c_2c_1 = 000$ are 3 ancilla ($c_0$ is not required), $a_0 = q_{move}$, $a_3a_2a_1 = 000$, $b_3b_2b_1b_0 = q_{head}^3q_{head}^2q_{head}^1q_{head}^0$ and $b_4$ is ignored.
The circuit in figure~\ref{qcirc:subt} shows the 4-bit modulo-12 quantum decrement circuit using the reverse order of blocks as the standard quantum adder.

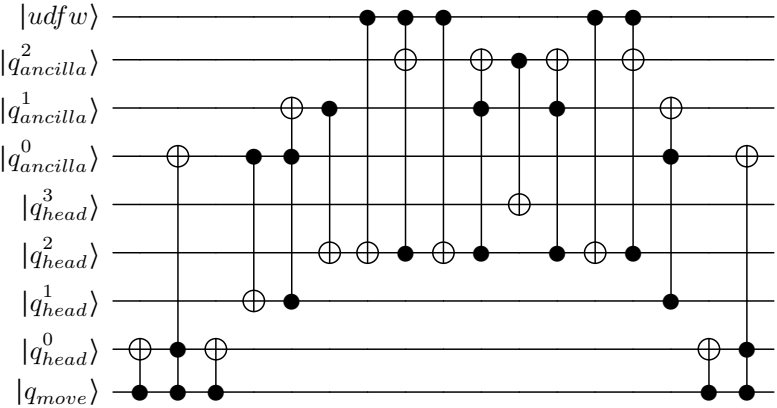
\begin{figure}[htb]
\centering 
\centerline{
\Qcircuit @C=0.6em @R=1.0em {
\lstick{\ket{udfw}}    			& \qw       & \qw       & \qw       & \qw       & \qw       & \qw       & \ctrl{5}  & \ctrl{1}  & \ctrl{5}  & \qw       & \qw       & \qw       & \ctrl{5}  & \ctrl{1}  & \qw       & \qw       & \qw       & \qw \\
\lstick{\ket{q_{ancilla}^2}}    & \qw       & \qw       & \qw       & \qw       & \qw       & \qw       & \qw       & \targ     & \qw       & \targ     & \ctrl{3}  & \targ     & \qw       & \targ     & \qw       & \qw       & \qw       & \qw \\
\lstick{\ket{q_{ancilla}^1}}    & \qw       & \qw       & \qw       & \qw       & \targ     & \ctrl{3}  & \qw       & \qw       & \qw       & \ctrl{-1} & \qw       & \ctrl{-1} & \qw       & \qw       & \targ     & \qw       & \qw       & \qw \\
\lstick{\ket{q_{ancilla}^0}}    & \qw       & \targ     & \qw       & \ctrl{3}  & \ctrl{-1} & \qw       & \qw       & \qw       & \qw       & \qw       & \qw       & \qw       & \qw       & \qw       & \ctrl{-1} & \qw       & \targ     & \qw \\
\lstick{\ket{q_{head}^3}}       & \qw       & \qw       & \qw       & \qw       & \qw       & \qw       & \qw       & \qw       & \qw       & \qw       & \targ     & \qw       & \qw       & \qw       & \qw       & \qw       & \qw       & \qw \\
\lstick{\ket{q_{head}^2}}       & \qw       & \qw       & \qw       & \qw       & \qw       & \targ     & \targ     & \ctrl{-4} & \targ     & \ctrl{-3} & \qw       & \ctrl{-3} & \targ     & \ctrl{-4} & \qw       & \qw       & \qw       & \qw \\
\lstick{\ket{q_{head}^1}}       & \qw       & \qw       & \qw       & \targ     & \ctrl{-3} & \qw       & \qw       & \qw       & \qw       & \qw       & \qw       & \qw       & \qw       & \qw       & \ctrl{-3} & \qw       & \qw       & \qw \\
\lstick{\ket{q_{head}^0}}       & \targ     & \ctrl{-4} & \targ     & \qw       & \qw       & \qw       & \qw       & \qw       & \qw       & \qw       & \qw       & \qw       & \qw       & \qw       & \qw       & \targ     & \ctrl{-4} & \qw \\
\lstick{\ket{q_{move}}}         & \ctrl{-1} & \ctrl{-1} & \ctrl{-1} & \qw       & \qw       & \qw       & \qw       & \qw       & \qw       & \qw       & \qw       & \qw       & \qw       & \qw       & \qw       & \ctrl{-1} & \ctrl{-1} & \qw \\
}
}
\caption{Modulo-12 decrement quantum circuit for QPULBA 2-2-1}
\label{qcirc:dec}
\end{figure}

The circuit can be simplified considering some of the control qubits are always in the $0$ state, so those CNOT/Toffoli gates can be ignored.
The optimized increment circuit for the QPULBA 2-2-1 is shown in figure~\ref{qcirc:dec}.

\newpage
\subsubsection{Reset} \label{s6s3s5}

\begin{figure}[ht]
    \centering % \captionsetup{justification=centering} % trim: LBRT
    \includegraphics[clip, trim=18cm 0cm 0cm 0cm, width=0.55\textwidth]{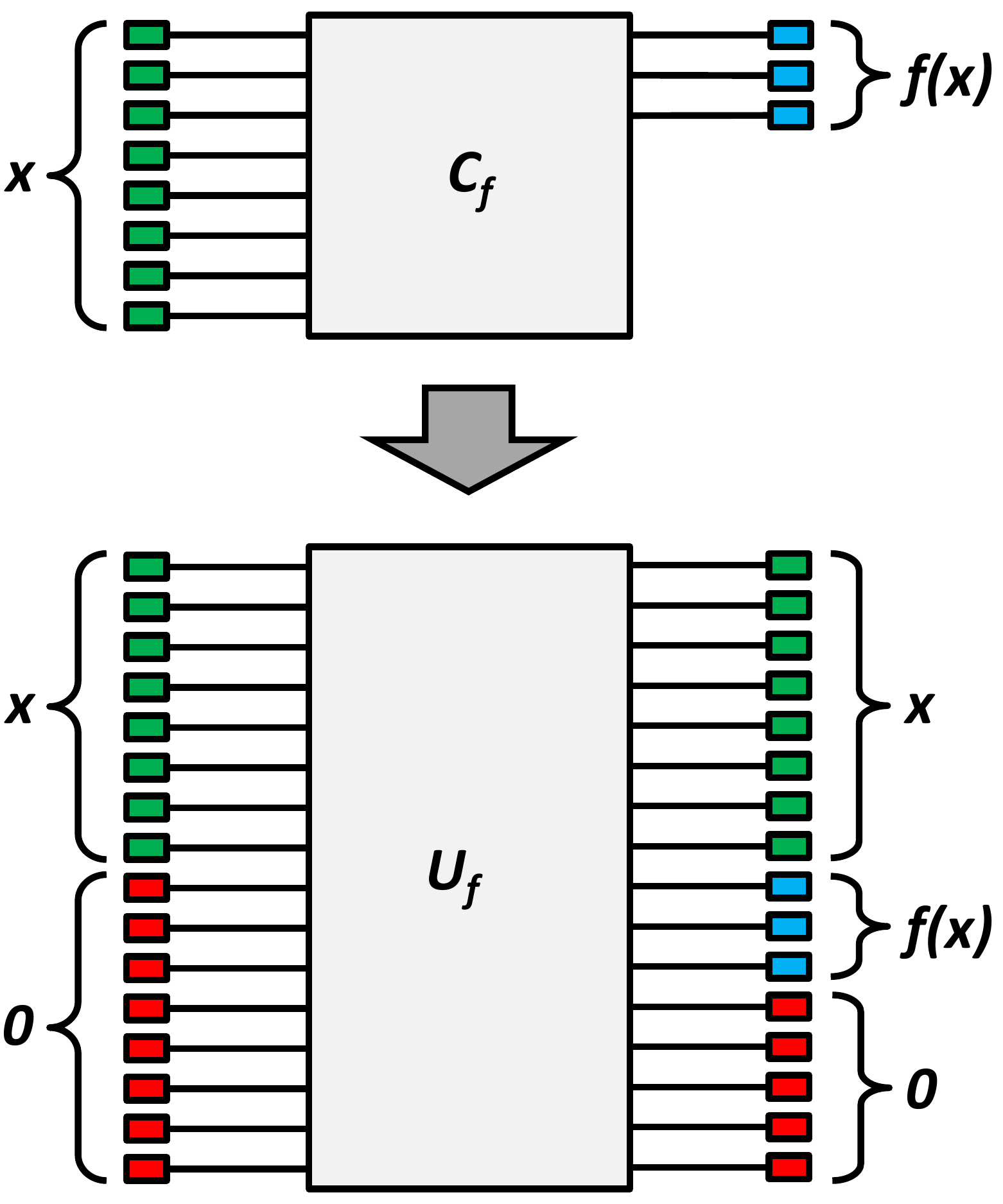}
    \caption{Reversible circuit compilation strategy}
    \label{fig:rev}
\end{figure}

Quantum logic is universal and can implement any classical Boolean logic function using only the Toffoli (CCNOT) or Fredkin (CSWAP) gate.
However, it is not always possible to uncompute all ancilla qubits.
Specifically, if the function to be implemented is irreversible, e.g. AND, extra qubits are needed to construct the reversible quantum circuit that preserves the unitary property.
The general strategy to compile a classical function to quantum logic is shown in figure~\ref{fig:rev}.

Other research has focused on the quantum circuit generation of the constrained subset of reversible automata.
However, for the QPULBA case, we intend to evolve a superposition of all possible classical functions/programs $C_f$ that can be represented by the description number encoding.
These functions include both reversible as well as irreversible functions, thus, we cannot uncompute away the computation history of the state transition.

Both the state and the read together preserve the evolution history.
Thus, we need ancilla qubits in each step of the computation that would hold the transition history for the QPULBA.
This limits the number of steps of the QPULBA we can implement or simulate.
Besides the state and read, the write and move qubits need to be reset in each cycle.
This is implemented by calling the FSM transition function once again with the previous state and the read.

% https://www.researchgate.net/publication/224161019_Reversing_deterministic_finite_state_machines

\newpage
\section{Implementation and simulation results} \label{s7}

In this section, we present the circuit implementation of QPULBA.
This was implemented on 2 different programming platforms, OpenQL~\cite{khammassi2020openql} developed at the Delft University of Technology and IBM's Qiskit~\cite{larose2019overview}.
Our copy-left AGPLv3 licensed implementation can be found on:
\\\href{https://github.com/Advanced-Research-Centre/QPULBA}{https://github.com/Advanced-Research-Centre/QPULBA}

We implemented 2 cases of QPULBA, with 1 and 2 states: the full circuit and cycle simulation of QPULBA 1-2-1, and 
a limited simulation of the units for QPULBA 2-2-1, as presented below.

\subsection{QPULBA 1-2-1}

Our implementation is scalable to any $m$-state $n$-symbol QPULBA.
The entire circuit for the 1-state 2-symbol case requires much less qubits, thus we were able to simulate it classically.
Note that there is no need to store the state anymore thereby reducing the qubit complexity greatly.
\begin{verbatim}
Number of 1-state 2-symbol 1-dimension QPULBA: 16

FSM     : [0, 1, 2, 3]
STATE   : []
MOVE    : [4]
HEAD    : [5, 6]
READ    : [7]
WRITE   : [8]
TAPE    : [9, 10, 11, 12]
ANCILLA : [13, 14, 15]
\end{verbatim}

\begin{figure}[hbt]
    \centering % \captionsetup{justification=centering} % trim: LBRT
    \includegraphics[clip, trim=22cm 0cm 0cm 9.4cm, width=0.55\textwidth]{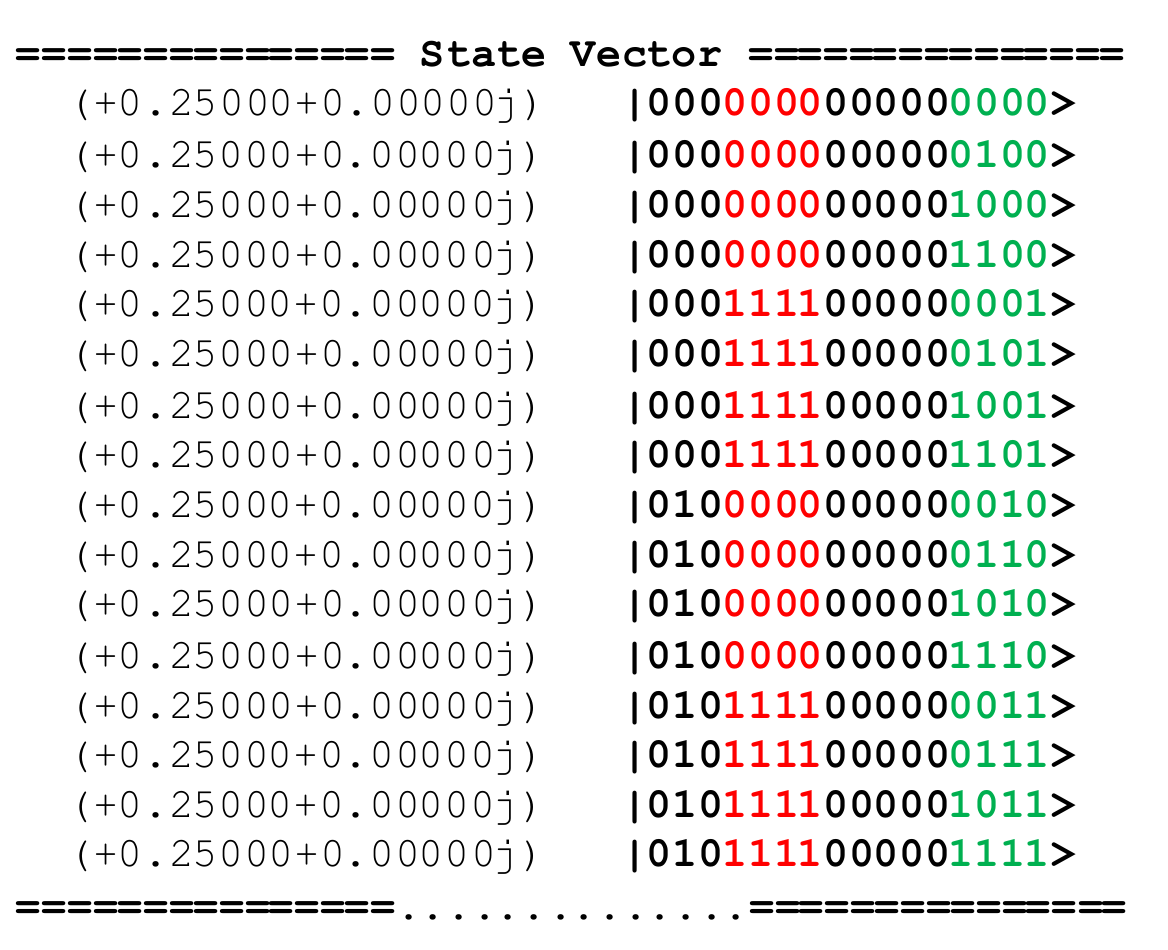}
    \caption{Test result for the QPULBA 1-2-1 showing the FSM description/program number (green) and output tape (red)}
    \label{fig:res121_isv}
\end{figure}

The full circuit was simulated for 4 cycles as discussed in \S~4(d) \ref{s5s4s3}.
The final state vector obtained after 4 cycles is shown in figure~\ref{fig:res121_isv}.
The FSM qubits encoding the description/program number (in green) and the output on the tape (in red) bit strings match with the classical enumeration in Table~\ref{tab:utm121}.
Thus, if we measure only the tape in the standard computational basis, we will obtain an equal statistical distribution of the $0000$ and $1111$ states.

\subsection{QPULBA 2-2-1}

The entire circuit for the 2-state 2-symbol case can be compiled from the parts described in \S~\ref{s6}.
This was implemented in a scalable manner on OpenQL and Qiskit.

\subsubsection{Full circuit compilation}

The qubit allocation for the full circuit, considering 1 cycle is:
\begin{verbatim}
Number of 2-state 2-symbol 1-dimension QPULBA: 4096

FSM     : [0, 1, 2, 3, 4, 5, 6, 7, 8, 9, 10, 11]
STATE   : [12, 13]
MOVE    : [14]
HEAD    : [15, 16, 17, 18]
READ    : [19]
WRITE   : [20]
TAPE    : [21, 22, 23, 24, 25, 26, 27, 28, 29, 30, 31, 32]
ANCILLA : [33, 34, 35]
\end{verbatim}
For each further cycles, we need 2 qubits to store the computation history.

The exact gate complexity depends on the considered primitive.
We use Hadamard, Pauli-X, CNOT, Toffoli and SWAP as our gate set.
Multi-qubit controlled-NOT gates are decomposed using the borrowed-ancilla strategy outlined in \cite{gidney_2018}.
One cycle of the QPULBA uses $627$ gates: $476$ Toffoli, $126$ Pauli-X, $12$ CNOT, $12$ Hadamard and $1$ SWAP gate.
The Qiskit circuit drawing and generated OpenQASM can be found on the repository.

\subsubsection{Unit tests}

While we were able to compile the full circuit, the large number of qubits limits classically simulating the circuit on our available hardware.
The exponential simulation complexity of quantum algorithms on classical hardware in terms of memory resource is indeed the driving factor for research on physical implementation of quantum accelerators.
To complement our design, we developed unit tests for each part of the mechanistic perspective.

\begin{figure}[hbt]
    \centering % \captionsetup{justification=centering} % trim: LBRT
    \includegraphics[clip, trim=0cm 0cm 0cm 2cm, width=\textwidth]{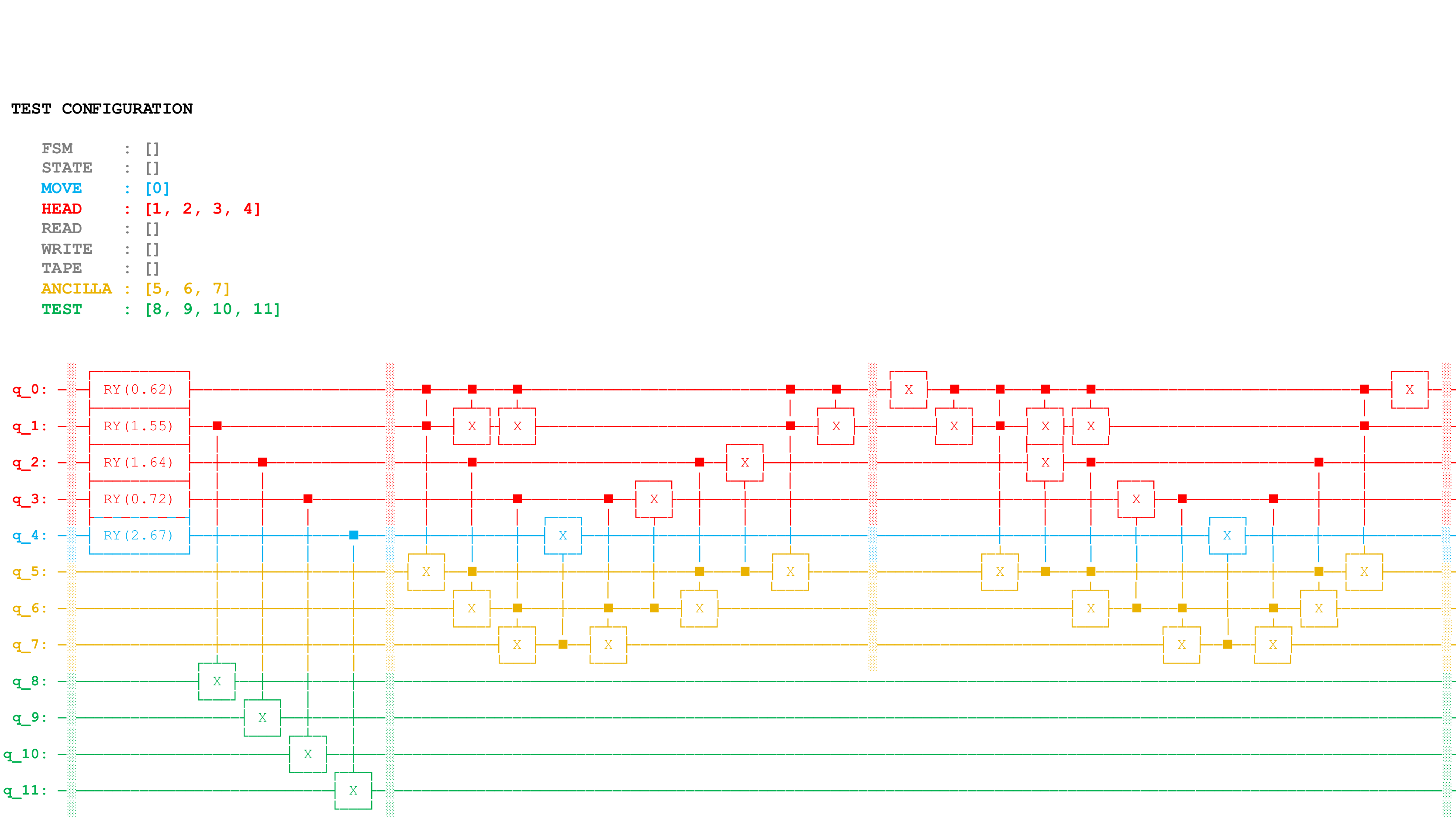}
    \caption{Test circuit for the QPULBA 2-2-1 move block}
    \label{fig:res_circ}
\end{figure}

We successfully tested each of the 6 parts of the QPULBA, i.e. initialize, read, FSM, write, move and reset.
The unit test simulations are tractable as each part concerns only a subset of the qubits.
The inputs are put into an unequal superposition of values using random rotations about the Y-axis (that maintains the amplitude in the real domain).
This allows us to individually inspect each basis state changes in contrast to an equal superposition using Hadamard gates.
For quantum acceleration of classical algorithms, the ZX-plane of the Bloch sphere is enough to take advantage of the destructive interference of quantum superpositions over classical probabilistic computing, thereby reducing the cost of classical simulation using complex representations of the amplitude.

To track the output changes, each qubit of the target register is entangled with a test register using CNOT gates.
Thus, the old value and the new value of the basis state's bit string can be inspected similar to an associative memory.

Here, as an example, we show the unit test circuit for the move step, which requires 8 qubits for the circuit and 4 addition qubits for testing.
The circuit is shown in figure~\ref{fig:res_circ}.
The initial part before the first barrier is the test configuration and the second part is of the move circuit.

\begin{figure}[hbt]
    \centering % \captionsetup{justification=centering} % trim: LBRT
    \includegraphics[clip, trim=22cm 0cm 0cm 1cm, width=0.55\textwidth]{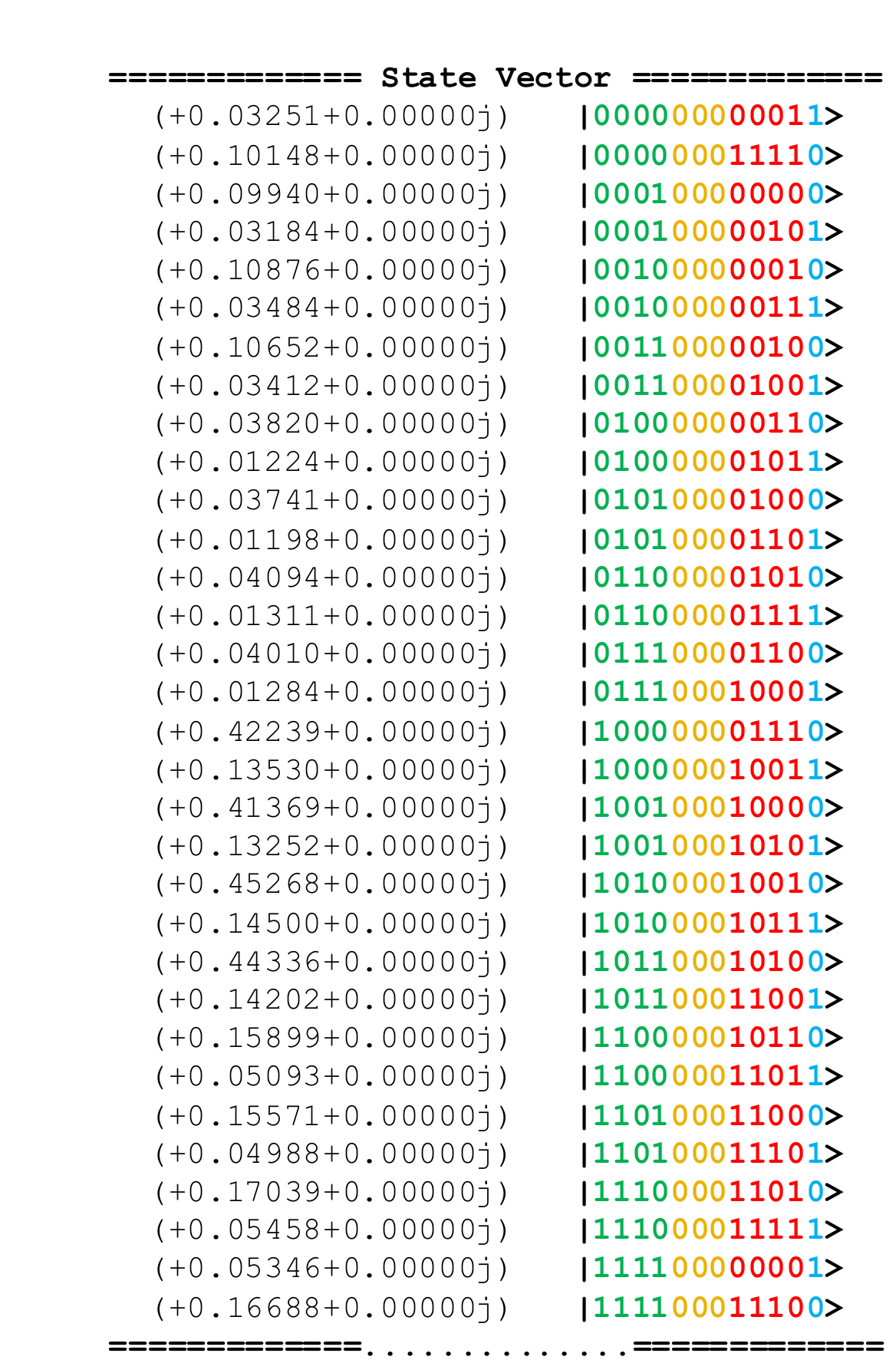}
    \caption{Test result for the QPULBA 2-2-1 move block showing the move (blue), head (green), ancilla (yellow) and test (green) qubits}
    \label{fig:res_isv}
\end{figure}

For the move circuit, the binary string of the qubits associated with the head state (in red) increments by 1 if the move qubit (in blue) is 1 and decrements by 1 otherwise.
This is verified by the internal state vector output of the simulation as shown in figure~\ref{fig:res_isv}.

% https://quantum-computing.ibm.com/docs/iqx/visualizations

\newpage
\section{Conclusion} \label{s8}

The mechanistic model of computation as exhibited by a Turing machine defines an algorithm as an initial input to final output transformation on the tape memory by a program defined as a finite state machine.
The set of transformations a computation model can undergo and the resulting space of outputs is central to understanding the causal structure of a physical phenomena for scientific modeling and hypothesis testing.
While it has many applications, except for the trivial cases, this remain intractable on classical computers.
This is because the space of all possible transformations grows exponentially with the number of states and symbols of the automata.

In this research, we explore the distinctive advantages for classical automata simulation offered by the alternate paradigm of quantum computation.
We complement the recently proposed~\cite{molina2019revisiting} circuit design of a quantum Turing machine from a mechanistic perspective with realistic assumptions on runtime and qubit resources.
In our design, we follow the computation model of a quantum parallel universal linear bounded automata.

We present the exact scalable circuit using standard quantum gates required to simulate a superposition of programs of this automata, thereby obtaining the distribution of their evolution after a predetermined number of cycles.
This algorithm can be readily ported on a quantum accelerator stack~\cite{bertels2020quantum} with any sufficiently advanced gate-model quantum computing hardware, in terms of qubits, connection topology and error rates.
We present our results of the implementation of two cases of the automata on two quantum programming platforms, OpenQL and Qiskit.
We simulated and verified 4 cycles of the 1-state 2-symbol quantum parallel universal linear bounded automata with 16 qubits and compiled and unit tested the 2-state variant that requires 36 qubits.

Quantum automata has promising advantages in extending the applications of approximating algorithmic metrics by enumerating automata configurations~\cite{zenil_2020}, as will be explored in our future research.
Applications in soft-computing, specifically accelerating the estimating of metrics like algorithmic probability~\cite{sarkar2021} and algorithmic complexity, are essential primitives in the fields of genomics, artificial life and artificial intelligence.